\documentclass{iopjournal}
\usepackage{lmodern}
\usepackage{natbib}
\usepackage{amsmath,amssymb}
\usepackage{ulem}
\usepackage{subfigure}

\begin{document}

\articletype{Topical Review} 

\title{Exploring Physics beyond the Standard Model from kHz-Gravitational-Wave Signals of Core-Collapse Supernovae}

\author{Kei Kotake$^1$\orcid{0000-0003-2456-6183}, Takami Kuroda$^2$\orcid{0000-0001-5168-6792}} 

\affil{$^1$Department of Applied Physics, Fukuoka University, Nanakuma, Jonan, 8-19-1, Fukuoka, 814-0180, Japan}

\affil{$^2$Max-Planck-Institut f\"{u}r Gravitationsphysik, Am M\"{u}hlenberg 1, D-14476 Potsdam-Golm, Germany}

\email{kkotake@fukuoka-u.ac.jp, takami.kuroda@aei.mpg.de}

\begin{abstract}
Recent advances in multidimensional modeling of core-collapse supernovae (CCSNe) have enabled detailed predictions of high-frequency gravitational-wave (GW) signals, offering a new probe of extreme matter and gravity.
A proto-neutron star (PNS) emits quasi-continuous GWs through the excitation of its characteristic oscillation modes.
State-of-the-art CCSN simulations show that these oscillations, in particular $g$- and $f$-modes, dominate the GW spectrum, with frequencies rising from a few hundred hertz to the kilohertz (kHz) range as the PNS contracts and its compactness increases in the post-bounce phase.
Therefore, the temporal evolution of these GW frequencies, if detected, would provide a direct and quantitative tracer of the PNS internal structure and the surrounding explosive dynamics.
In addition to such a GW emission mechanism associated with PNS contraction, which is currently thought to be the {\it standard} GW emission mechanism, fully general relativistic (GR) simulations have revealed additional GW sources linked to more exotic physical processes.
In highly massive progenitors, continuous mass accretion drives rapid PNS contraction and early black-hole (BH) formation, producing strong kHz GW emission that abruptly ceases when the PNS core is swallowed by the BH horizon.
Similarly, a strong first-order quantum chromodynamics (QCD) phase transition can induce a secondary collapse and rebound of the nascent quark core, generating powerful, millisecond-duration GW bursts with frequencies exceeding $\sim$2\,kHz.
Alternative theories of gravity, such as scalar-tensor frameworks, predict spontaneous scalarization that can trigger multiple collapses of the PNS, yielding analogous high-frequency and broadband GW signals.
The combined analysis of these GW signals, together with their detection by next-generation GW detectors, offers a promising multi-messenger pathway to identify smoking-gun signatures of new physics beyond the standard model of the CCSN GW mechanism and general relativity.
\end{abstract}

\section{Introduction}
\label{sec:Introduction}
High-frequency GWs emitted from CCSNe offer valuable insights into the early evolution of newly formed PNSs.
Massive stars with initial masses $\gtrsim8\,{\rm M}_\odot$ undergo gravitational collapse of their iron cores once the core mass approaches the Chandrasekhar limit and form a PNS.
This moment marks an important early phase in the evolution of compact stellar remnants and their subsequent diverse evolutionary paths \cite[see][for a recent review]{Janka25_review}, and is accompanied by the emergence of characteristic multi-messenger signals such as GWs and neutrinos.

From the perspective of GW emission, the PNS evolution plays the most crucial role.
After its formation, the PNS continues to accrete matter from its surroundings.
This ongoing accretion increases the compactness of the PNS ($M/R$, with $M$ and $R$ being the mass and radius of the PNS, respectively).
Such a compact remnant strongly distorts the surrounding spacetime, and thus governs the dynamics there.
One important consequence of continued accretion is the excitation of the PNS's intrinsic oscillation modes \citep{Kokkotas99}, such as the fundamental ($f$-) and gravity ($g$-)mode oscillations \cite{jim15,Torres-Forne19,Morozova18,marie_anne21,bruel23}.
Excitation of these oscillation modes is the canonical mechanism for {\it standard} GW production in CCSNe.
Besides these GWs arising from oscillating PNSs, progenitor rotation is also a promising source of high-frequency GWs, from which we can infer the rotational properties of the PNS at bounce, and even those of the progenitor star itself. Previous studies \citep{Dimmelmeier08,Scheidegger10,Ott11,Abdikamalov14} have shown that rotational core bounce creates a GW burst, commonly referred to as a Type I GW signal \citep{Zwerger97,Dimmelmeier02B}, with an emission band covering a wide frequency range from $\sim100$ to $\lesssim1000$~Hz. Ref.~\cite{Dimmelmeier08} clearly indicated that the peak GW frequency originating from rotating PNSs tends to cluster predominantly depending on the strength of the precollapse rotation rate.
In the post-bounce phase, although PNS rotation can sometimes produce characteristic GW signals through, for example, non-axisymmetric rotational instabilities or spiral standing accretion shock instabilities \citep{Blondin07_nat}, and these phenomena are certainly worth detailed investigation, the overall frequency range generally falls within the low to moderate regime, at most
$\lesssim$ several hundred Hz\footnote{We should note that rotating models can indeed generate high-frequency GWs ($\sim1000$~Hz) \citep{Scheidegger10,KurodaT14,Takiwaki18,Shibagaki21,bugli23,sykes26}. However, we consider that these high-frequency components mainly originate from PNS oscillations themselves, and that rotational effects may further shift the frequency by, at most, a few hundred Hz, depending on the degree of rotation (see main text).}. This limitation arises because such {\it global} non-axisymmetric rotational instabilities have characteristic spatial scales of
$\sim10^7$~cm \citep{Blondin07_nat,Scheidegger10,KurodaT14,Takiwaki18,Shibagaki21}, resulting in characteristic frequencies that are constrained to at most several hundred Hz by the sound speed or the angular velocity of matter (both typically below $\sim10^9$~cm~s$^{-1}$).
In this review, we thus choose not to present a detailed discussion of the rotational impact on {\it high-frequency} GWs (see \cite{bernhard2026} for a recent review).

Importantly, PNS oscillations can be continuously excited as long as the accreted material retains sufficient momentum to penetrate into the PNS surface \citep{Murphy09,BMuller13}.
This makes the resulting GWs long-lasting signals, occasionally persisting for several seconds after bounce.
As another key feature, the PNS becomes more compact over time and the frequencies of the excited modes increase accordingly, reflecting changes in the internal structure of the PNS \cite{EMuller97,EMuller04,EMuller12}.
Although the typical oscillation frequencies differ from mode to mode, current full-fledged CCSN simulations (in different contexts) indicate that they reach the $\sim$kHz range within approximately one second after PNS formation \citep{Andresen17,radice19,Powell20,Andresen21,eggenberger21, Takiwaki21,colter22,Nakamura22,Vartanyan23,Mezzacappa23,tony2024,choi24,ehring26,Lella26}. Here, by ``full-fledged'' CCSN models we refer, at minimum, to simulations that solve multi-energy neutrino radiation transport coupled to full-scale multi-dimensional hydrodynamics, including state-of-the-art neutrino opacities and general relativistic (GR) gravity (see \cite{Janka25_review,tony_review,yamada_review,Burrows21_review,Kotake12_ptep} for reviews).
These facts imply that the detection of such high-frequency GWs, and their temporal modulation due to the quasi-hydrostatically contracting PNS, can provide live information on the time evolution of the PNS core, namely the formation physics of a neutron star (NS).

In contrast to this quasi-static evolution of the PNS and its associated standard GW signals, more dynamical processes have recently attracted increasing attention from the perspectives of both new explosion scenarios and {\it beyond-the-standard} GW emission mechanisms.
A prominent example arises from a QCD phase transition \citep{Sagert09,Nakazato13,Fischer18}, where hadronic matter converts into quark matter inside the PNS.
Assuming a strong first-order phase transition \cite{takahara,gentile}, the equation of state (EoS) suddenly softens at the interface between the hadronic and quark phases and the core undergoes a {\it second} collapse.
Thereafter the second bounce---this time of the quark core---occurs and produces powerful shock waves, whose explosion energy and velocity reach $\sim10^{51}$\,erg and $\sim0.1c$, respectively, in spherically symmetric (1D) simulations \citep{Fischer18,Andia23} for moderately massive progenitor stars.
Or, if the progenitor star is too massive, the second collapse directly results in BH formation \citep{KurodaT22}.
As remarkable multi-dimensional effects, these second shocks excite strong convective motions along their passage, from which burst-like GWs with $\sim$kHz frequencies can be emitted \citep{Zha20,KurodaT22,jakobus_2023}.
In addition to the QCD phase transition, it has been suggested that spontaneous scalarization allowed in certain ST (scalar tensor) theories of GR can also induce a similar second collapse of the PNS, resulting in the emission of high-frequency GWs as well \citep{KurodaT23STT}.

BH formation is also regarded as a crucial source of kHz GWs.
The most recent fully relativistic BH formation models \citep{Powell21,Rahman2022,KurodaT23BH} show kHz GW emission prior to BH formation.
The underlying emission mechanism is essentially the same as in canonical CCSN models whose remnants are expected to be NSs, namely the excitation of PNS oscillations.
The major difference stems from the presence of a highly compact---and, more importantly, rapidly contracting---PNS resulting from the more intense mass accretion characteristic of very massive progenitors.
These conditions lead to a rapidly increasing frequency of various intrinsic PNS oscillations and of the associated GW peak frequency.
For massive progenitors heavier than $\sim40\,{\rm M}_\odot$, it has been suggested that BH formation occurs relatively early in the post-bounce phase, i.e., several hundred milliseconds after bounce \citep{Liebendorfer04,Sumiyoshi07,O'Connor11,Chan&Muller18,KurodaT18}.
Therefore, the GW signatures are expected to be significantly different from those of standard GW emission mechanism, in which it takes about one second until the GW frequency reaches $\sim$kHz.
This fact supports the idea that the early detection of such high-frequency GWs can provide direct evidence of BH formation, otherwise inaccessible to us because it is hidden under the thick envelopes of massive stars.

These recent findings suggest that understanding both standard and beyond-the-standard high-frequency GW emission mechanisms can greatly enhance our ability to interpret future observations, advance our knowledge of matter in extreme states, and help to constrain various CCSN scenarios.
In this review, we discuss a variety of GW emission mechanisms obtained by the most advanced numerical modeling of CCSNe, with a particular focus on the high-frequency regime in and beyond the standard GW emission model, which still remains poorly explored.
Our presentation will proceed as follows: we begin by explaining the standard GW emission mechanism in Section~\ref{sec:Standard GW emission mechanism}.
Section~\ref{sec:BH formation} discusses the case with BH formation.
Section~\ref{sec:Exotic star formation} is devoted to recent progress in more exotic scenarios, where stronger connections with other disciplines, such as nuclear physics and gravitational theory, are more prominent.
We give a summary in Section~\ref{sec:Conclusion}.

\section{High frequency GWs from the standard GW emission mechanism: PNS contraction}
\label{sec:Standard GW emission mechanism}

In this section, we give a brief overview of the GW emission mechanisms originating from the oscillating PNS, based on the most up-to-date CCSN GW predictions.
At present, these GW signatures are regarded as {\it standard} (see \cite{Ott09,fryer11,Kotake13,ernazar2020,bernhard2026} for reviews).
For clarity, we focus on the following questions: what is the actual source of these standard GWs, what triggers their emission, and, if detected, what can we learn from them?

\subsection{GWs from oscillating PNS}
\label{GWs from oscillating PNS}
We begin our discussion of the standard GW emission mechanism using results from a three-dimensional (3D) CCSN model \cite{Nakamura22}.
This model employed a binary stellar evolutionary model of SN~1987A \cite{urushibata} and resulted in a successful explosion.
\begin{figure*}[t!]
\centering
\includegraphics[width=0.8\columnwidth,angle=0]{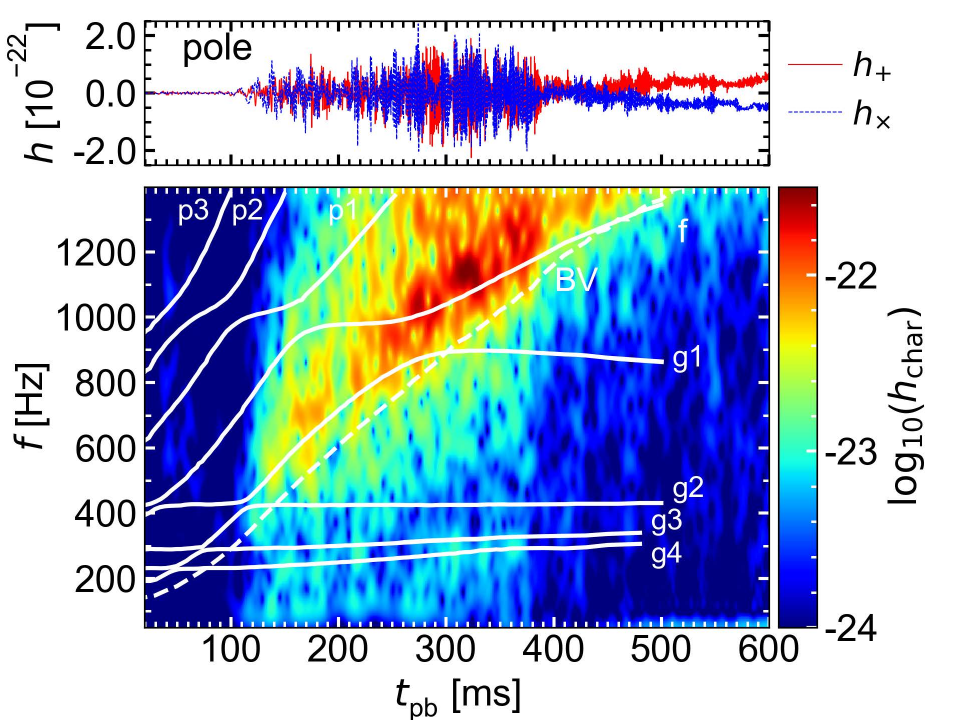}
\caption{GW waveform (top panel) and the spectrogram from a 3D CCSN model (bottom panel) using a binary progenitor model of SN~1987A \cite{Nakamura22}. An observer is assumed to be located along the equatorial direction at a distance of 10~kpc.
In the bottom panel, some characteristic GW frequencies of the $f$, $g_i$, and $p_i$ modes with the subscript of $i$ being the radial node numbers are overplotted, where an open source code (GREAT; \cite{Torres-Forne19}) was used to perform the eigenmode analysis. The dashed white line labeled as "BV" represents the Brunt–V\"ais\"al\"a frequency (equivalently, the peak GW frequency (Equation (1), see text). This figure is taken from \cite{Nakamura22}, nonetheless slightly modified.
\label{fig:Nakamura22}}
\end{figure*}
Fig.~\ref{fig:Nakamura22} shows the GW waveform of the plus and cross modes ($h_+$ and $h_\times$, top panel) and a colour map of the frequency components $f$ (spectrogram, bottom panel) as a function of time after bounce ($t_{\rm pb}$), assuming that an observer is located along the $x$-axis of the simulation coordinates (i.e. at the equator) at the Galactic Centre at a distance of 10~kpc.
One can identify a typical time series of CCSN GW features in both polarizations.
First, during the initial $\sim70$~ms after bounce, a GW signal appears across a wide frequency range, followed by a relatively quiet phase from $t_{\rm pb}\sim70$ to $\sim150$~ms.
Subsequently, non-linear fluid motions behind the shock excite strong high-frequency signals, starting at $f\sim500$~Hz around $t_{\rm pb}\sim150$~ms and increasing to $f>1000$~Hz by $t_{\rm pb}\sim400$~ms.
As the shock expands and the non-linear motions behind it are diluted ($t_{\rm pb}>400$~ms), the amplitude of these signals gradually decreases.
In addition, another component appears at around 100–300~Hz after $t_{\rm pb}\sim150$~ms.
For later reference, we briefly summarize the physical origin of these {\it standard} GW features.

The core bounce marks the birth of the PNS, at whose surface a strong shock is generated by the repulsive nuclear force.
However, the bounce shock stalls in the core at about $t_{\rm pb}\sim10$~ms because of photo-disintegration of the iron core and neutrino cooling behind the shock, and turns into a standing shock (see \cite{langanke03,Kotake06,Fischer14} for reviews). 
The rapid sequence of these events, occurring within this short time interval, leads to the so-called prompt convection ($0 \lesssim t_{\rm pb} \lesssim 70$~ms in Fig.~\ref{fig:Nakamura22}, top panel), where a characteristic entropy structure forms behind the shock: a convectively stable layer (with a positive entropy gradient) in the inner region $r\lesssim30$~km, and an unstable layer (with a negative entropy gradient) at larger radii near $r_{\rm shock}$ \citep{Liebendorfer01,Buras06a}.
The prompt convection generates GW emission within the first several tens of ms after bounce (e.g. top panel of Fig.~\ref{fig:Nakamura22}) in the unstable inner region.
Later on, after a quiescent phase ($70 \lesssim t_{\rm pb} \lesssim 120$~ms in Fig.~\ref{fig:Nakamura22}), convective plumes penetrate vigorously into the underlying convectively stable layer, which becomes the main driver of oscillations of the PNS core and hence of GW emission. Furthermore, the standing-accretion-shock-instability (SASI \cite{Foglizzo09,Foglizzo07,Foglizzo12,Foglizzo15}) also plays a pivotal role, depending upon the progenitor masses, in determining the gravitational waveform in the low-frequency ($\lesssim 200$ Hz) regime \cite{KurodaT16ApJL,Hayama18,Kawahara18,Mezzacappa23}.

A key insight into the properties of these GW signals is that their characteristic frequencies can be naturally explained by the Brunt–V\"ais\"al\"a frequency at the PNS surface (see the white dashed line in the bottom panel of Figure 1 labeled as "BV" \citep{Murphy09,BMuller13}, where the downflows of convective plumes are decelerated and redirected into upflows by the buoyancy force, i.e. they excite the $g$-mode.
This feature is common in many simulations \citep{Murphy09,BMuller13,KurodaT16,Torres-Forne18,Morozova18,Powell21,Nakamura22}, particularly in the early post-bounce phase ($t_{\rm pb}\lesssim500$~ms).
It is still not fully understood whether this initial $g$-mode phase continues throughout the subsequent PNS contraction phase \citep{Torres-Forne19} or smoothly transits to the fundamental ($f$-)mode \citep{Morozova18,Murphy25}, as suggested by Fig.~\ref{fig:Nakamura22}. 
Furthermore, localization of spatial origins of oscillations in the early postbounce phase is another intriguing topic.
We can expect, in a simplified picture, that there are three distinguishable spatial origins of GW emission: (i) PNS convection region appearing at $r\sim20-40$~km, stemming mainly from the negative lepton number gradient \citep{Bruenn04,Buras06b}, (ii) post-shock convection region at $r\sim100$~km driven by the negative entropy gradient \citep{Foglizzo06}, and (iii) the entire PNS itself, with the latter being responsible for the $f$-mode oscillation.
Although it is quite challenging to distinguish whether the GW signals originate from the global oscillation of the PNS (i.e., $f$-mode) or from more localized oscillations in the aforementioned two distinct convection regions, discrimination between the inner PNS convection region and the outer post-shock convection region appears to be relatively straightforward based on the GW spectrogram.
In general, the inner PNS convection region produces several times higher buoyancy frequency ($\gtrsim500$~Hz) than the outer post-shock convection region, whose typical frequency range appears at $\sim100$-200~Hz \citep{Murphy09}.
According to many previous SN GW modeling studies \citep{Murphy09,BMuller13,KurodaT16ApJL,Powell21,Nakamura22}, the initial GW spectrogram (focusing on the immediate post bounce phase $t_{\rm pb}\lesssim50$~ms) presents a wide frequency band ranging from $\sim100$ to $\lesssim1000$~Hz, implying that both PNS and post-shock convection regions contribute to the early GW spectrogram via convective overturns characteristic of each region.
However, the higher frequency band becomes less visible in the GW spectrogram \citep{Murphy09,BMuller13,KurodaT16ApJL,Powell21,Nakamura22} within the first $\sim100$~ms.
We therefore interpret the relatively low initial GW frequency ($\sim100$~Hz) as an indication that the major emission region is subsequently localized in the outer convection zone, where the emission can be sustained primarily by convective plumes accreting from above.

In any case, the ramp-up GW pattern clearly encodes information about how the PNS properties evolve with time after bounce and how strongly the PNS is hit by the surrounding convective plumes.
Importantly, the Brunt–V\"ais\"al\"a frequency and the $f$-mode depend on the surface gravity and the mean density of the PNS, respectively, and are therefore sensitive to the evolving mass and radius of the PNS, i.e. to the compactness parameter $M/R$.
As the supernova proceeds, continuous mass accretion leads to a gradual increase in the PNS compactness, which also affects the thermal evolution and the buoyancy profile.
This is reflected in the time evolution of the $g/f$-mode frequencies, which increase as the PNS contracts (see, e.g., \cite{Alfe23,Kawahara18} for the detection prospects of the $f/g$ modes).
Meanwhile, the amplitude of the perturbations—here mainly set by the mass accretion rate—largely determines the GW amplitudes.
Thus, detecting these ramp-up high-frequency GWs in the post-bounce phase would provide a clear probe of the changing internal structure and dynamics around the newly born PNS (or even a BH, as discussed below).

Recent full-fledged CCSN simulations indicate that typical GW frequencies appear at around a few $100$~Hz shortly after bounce ($t_{\rm pb}\lesssim50$~ms), although the exact value depends on the dominant mode excited and on other properties such as the progenitor mass, rotation, and the EoS.
Later, as the PNS contracts, these frequencies increase with time and reach the $\sim$kHz range within approximately one second after PNS formation \citep{Powell20,Nakamura22,Vartanyan23,Mezzacappa23,Lella26}.
The overall trend of this evolution in the GW spectrogram can be well fitted by an analytical expression for $f_{\rm peak}$, first proposed by \cite{Murphy09} and later refined by \cite{BMuller13} with GR corrections.
Here, $f_{\rm peak}$ describes the GW peak frequency in terms of the local Brunt–V\"ais\"al\"a frequency (the $g$-mode) and represents the inverse of the deceleration time scale of convective plumes impacting the PNS surface.
The restoring force of the $g$-mode acting on perturbed fluid elements depends on the local gravity, e.g. $\propto M_{\rm PNS}/R_{\rm PNS}^2$ at the PNS surface, where $M_{\rm PNS}$ and $R_{\rm PNS}$ are the PNS mass and radius, respectively.
As a result, $f_{\rm peak}$ scales as
\begin{equation}
    f_{\rm peak}\propto \frac{M_{\rm PNS}}{R_{\rm PNS}^2}.
\end{equation}
Note that this formula is known not only to capture well firstly the low $g$-mode frequency in the postbounce phase, but also the $f$-mode frequency up to the late postbounce phase \cite{sotani19,sotani_rev} (see also the white dashed line labeled as "BV" in Figure 1).
During the contraction phase, the PNS mass typically increases by several tens of percent, as long as it does not collapse into a BH.
In contrast, its radius shrinks by a factor of a few, from an initial size of $\sim50$–60~km down to $\sim10$~km.
Since the peak frequency scales with $M_{\rm PNS}/R_{\rm PNS}^2$, its evolution is mainly controlled by the contracting PNS radius \citep[see Fig.~16 of][for a quantitative discussion]{BMuller13}.
Because the initial GW peak frequencies are around $\sim100$~Hz during the early convection phase, the time when the frequency reaches the kHz range roughly corresponds to the NS radius having decreased to about one third of its original value.
A crucial point here is that the growth rates of the PNS mass and radius depend strongly on the progenitor properties (e.g. mass and spin), the nuclear EoS, and the explosion dynamics \cite{sotani_eos}.
By exploiting this dependence, future observations of the time evolution of high-frequency GWs are expected to place useful constraints on supernova theory and nuclear physics.

\subsection{Excited eigen modes}
\label{sec:Excited eigen modes}
A deeper understanding of these GWs requires not only accurate GW predictions from full-fledged CCSN simulations, which capture the non-linear dynamics, but also a clear picture of how PNS oscillations are excited from a more theoretical point of view.
For this latter approach, asteroseismology based on linear perturbation theory is particularly powerful, and Refs.~\cite{Torres-Forne18,Morozova18,Torres-Forne19} have made important contributions in this direction.
Assuming a quasi-hydrostatic configuration at each time slice, one solves an eigenvalue problem to obtain the eigenfrequencies of various oscillation modes inside the PNS.
These studies show that the dominant part of the long-lasting (beyond one second) GW signal from PNS oscillations is either a $g$-mode \citep{Torres-Forne19} or a $g$-mode that later changes smoothly into the fundamental ($f$-)mode \citep{Morozova18}.
A key difference between these two modes is the parameter that mainly sets their eigenfrequency.
While the $f$-mode depends mainly on the square root of the average PNS density (i.e. $\sqrt{M_{\rm  PNS}/R_{\rm PNS}^3}$), the $g$-mode depends on the surface gravity, $M_{\rm  PNS}/R_{\rm PNS}^2$.
Thus, for a given time evolution of PNS mass and radius, the frequencies of the different modes differ in both magnitude and time dependence.
However, a common finding is that there is a universal relation for the time evolution of these frequencies as a function of post-bounce time, which is rather insensitive to the progenitor mass and nuclear EoS at a qualitative level (e.g. \cite{sotani_eos}), and in which the PNS contraction (i.e. increasing $M_{\rm  PNS}/R_{\rm PNS}$) is the key pushing the oscillation frequencies into the kHz range.

Ref.~\cite{Nakamura22} also studied the time–frequency behaviour of perturbative fluid motions using the open-source General Relativistic Eigenmode Analysis Tool (GREAT; \cite{Torres-Forne19}), varying the boundary conditions as described in \cite{Sotani&Takiwaki20}.
Several characteristic modes were identified and overplotted on the GW spectrogram in the bottom panel of Fig.~\ref{fig:Nakamura22}.
The most prominent frequency component appears at approximately 150~ms after bounce, starting around 500~Hz and evolving beyond 1~kHz.
This behaviour can be well fitted by the $g_1$-mode at early times, which then shows a transition to the $f$-mode at about $t_{\rm pb}\simeq250$~ms \citep{Sotani17,Morozova18}.
We shortly mention that the deviation between the characteristic frequency obtained from a perturbation theory and the actual GW spectral peak of the order of $\sim100$~Hz seen at a late phase ($t_{\rm pb}\gtrsim400$~Hz) may arise from several uncertain factors in the perturbation analysis including the treatment of boundary conditions \cite{sotani19},  approximations of general relativity \cite{Sotani&Takiwaki20}, or simply the limitation of the {\it linear} perturbation approach itself (see \cite{sotani_rev} for a review).
In fact, these state-of-the-art CCSN simulations support the picture that the main GW component originates from either $g$- or $f$-mode oscillations of the PNS.

\subsection{Dependence on the progenitor mass}
\label{Dependence on the progenitor mass}
In the previous section, we discussed the general role of a contracting PNS and its surrounding convection in shaping the GW signal, especially the characteristic frequencies linked to $g/f$-mode oscillations.
In this section, we turn to the progenitor mass, which is one of the largest uncertainties in supernova modelling and is expected to affect both the overall PNS evolution and the standard GW emission process described above.
This is of particular interest because future GW observations of CCSNe may constrain progenitor models and the underlying explosion dynamics.

As mentioned above, mass accretion is the main factor that determines how the PNS mass and radius evolve.
It also acts as an external force that excites various oscillation modes inside the PNS.
The mass accretion history largely reflects the progenitor structure, especially its density profile.
In addition, the explosion dynamics are likely the next most important factor, because once an explosion sets in, the accretion onto the PNS and thus the GW emission tend to decrease.
Therefore, the emitted GWs are expected to encode direct information about the SN central engine.
To explore the impact of progenitor properties, especially the mass, and of explosion dynamics on the GW signal, we performed two-dimensional (2D) axisymmetric simulations of the core collapse of massive stars with four different zero-age-main-sequence (ZAMS) masses: $9.6$\,M$_\odot$ (hereafter labelled z9.6), $11.2$\,M$_\odot$ (s11.2), $50$\,M$_\odot$ (s50), and $70$\,M$_\odot$ (z70) \citep[see][for details of each progenitor model]{KurodaT22}, in full GR.
The supernova model is based on numerical relativity and solves the general relativistic (GR) neutrino radiation-hydrodynamics equations with a two-moment (M1) neutrino transport scheme \citep[details can be found in][]{KurodaT16}.
\begin{figure*}[t!]
\centering
\includegraphics[width=0.43\columnwidth,angle=0]{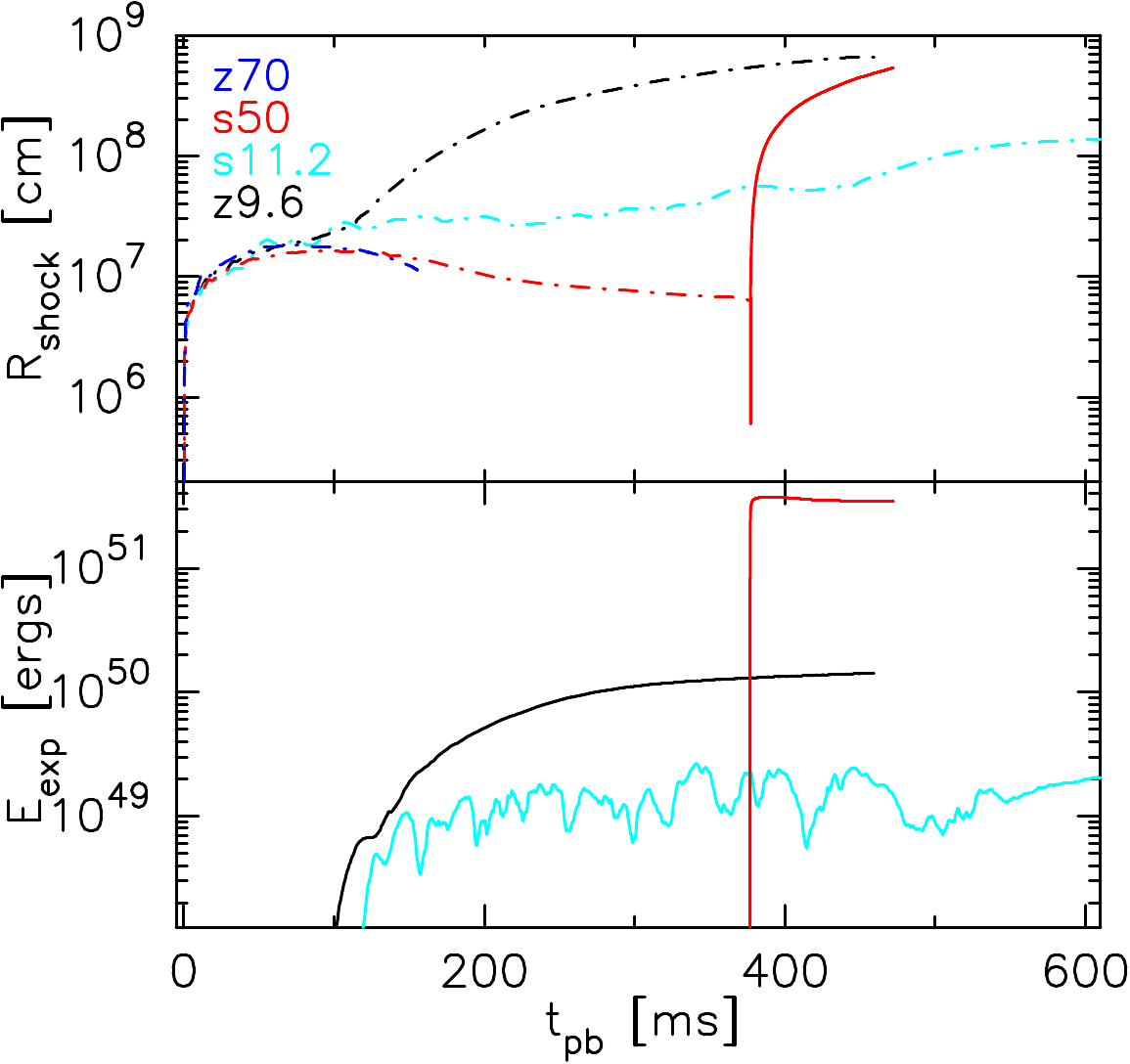}
\includegraphics[width=0.47\columnwidth,angle=0]{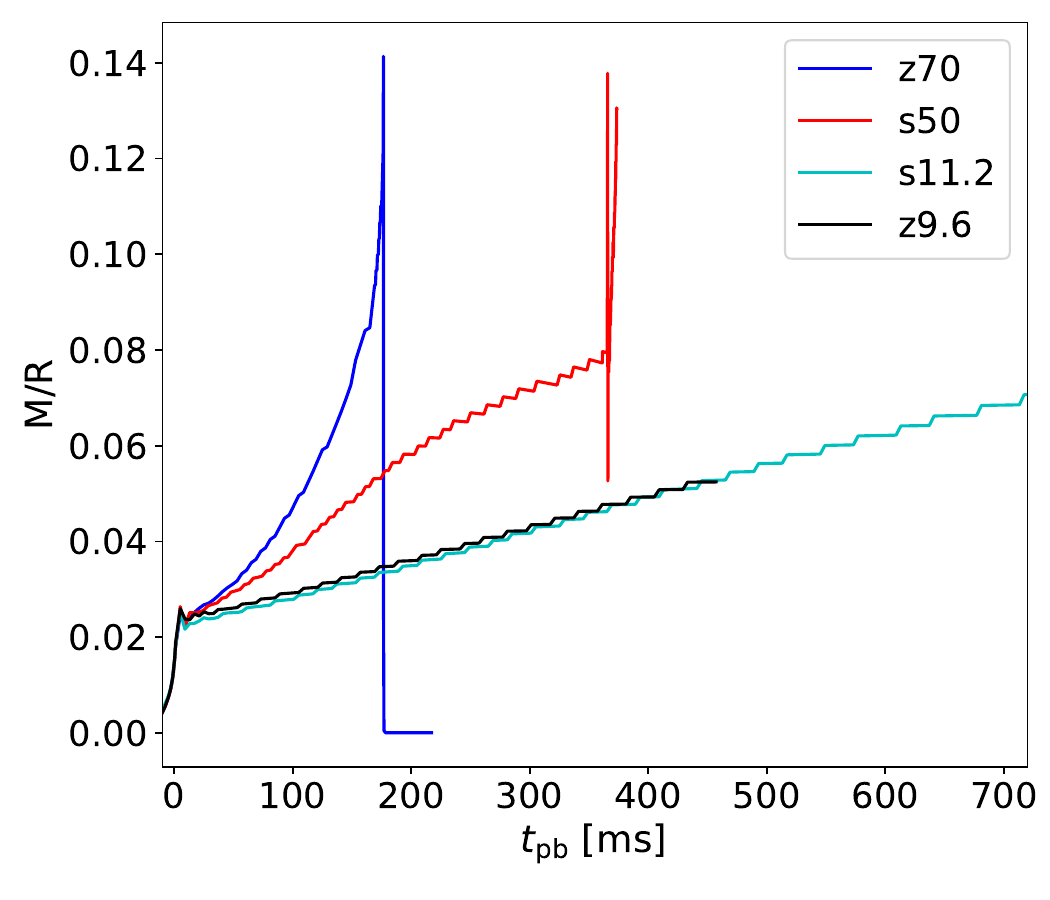}
\caption{Post-bounce evolution of the averaged shock radius $r_{\rm shock}$ (top left) and the diagnostic explosion energy $E_{\rm exp}$ (bottom left), taken from \citep{KurodaT22}, and the non-dimensional compactness parameter $M/R$ of the PNS for all models (right).
In the upper-left panel, we distinguish the core-bounce shock (dash-dotted line) and the second-bounce shock (solid line).
\label{fig:Evolution_Kuroda22}}
\end{figure*}
In the work,  an EoS including a first-order phase transition from hadronic matter to quark–gluon plasma \citep[for details, see][]{Fischer18,Bastian:2021} and up-to-date neutrino opacities \cite{Kotake18} were employed.

We briefly summarize the overall dynamics of all the computed models, because the explosion dynamics, whether the explosion succeeds or fails, leaves clear imprints in the GW signals. 
Fig.~\ref{fig:Evolution_Kuroda22} shows a wide variety of post-bounce evolution for different progenitor masses, where
the left panels show the shock radius (top) and diagnostic explosion energy (bottom), and the right panel shows the PNS compactness parameter for the four models.
The overall behaviour can be summarized as follows:
(1) the lower-mass models (z9.6 and s11.2) exhibit prompt-like explosions with modest explosion energies of $\sim10^{49-50}$\,erg, whereas the most massive model (z70) shows no shock revival and instead forms a BH at $t_{\rm pb}=177$~ms;
(2) the evolution of the compactness parameter $M/R$ correlates with the progenitor mass, with a larger progenitor mass yielding a higher compactness and a faster growth of $M/R$, although this statement is based on a limited set of models.
In the following, we mainly discuss how the PNS compactness ($M/R$) and the mass accretion rate affect the GW emission, focusing first on the two successful explosion models z9.6 and s11.2.
These two models may be viewed as representative examples of standard CCSN and GW emission mechanisms.
For more details on the dynamics associated with BH formation and the phase transition from hadronic to quark matter, we refer the reader to \cite{KurodaT22,KurodaT23BH}.
High-frequency GW emission in the context of such beyond-the-standard GW emission mechanisms will be discussed in Sections~\ref{sec:BH formation} and \ref{sec:Exotic star formation}.

\begin{figure*}[htbp]
\begin{center}
\subfigure[]{\includegraphics[width=0.7\columnwidth,angle=0.]{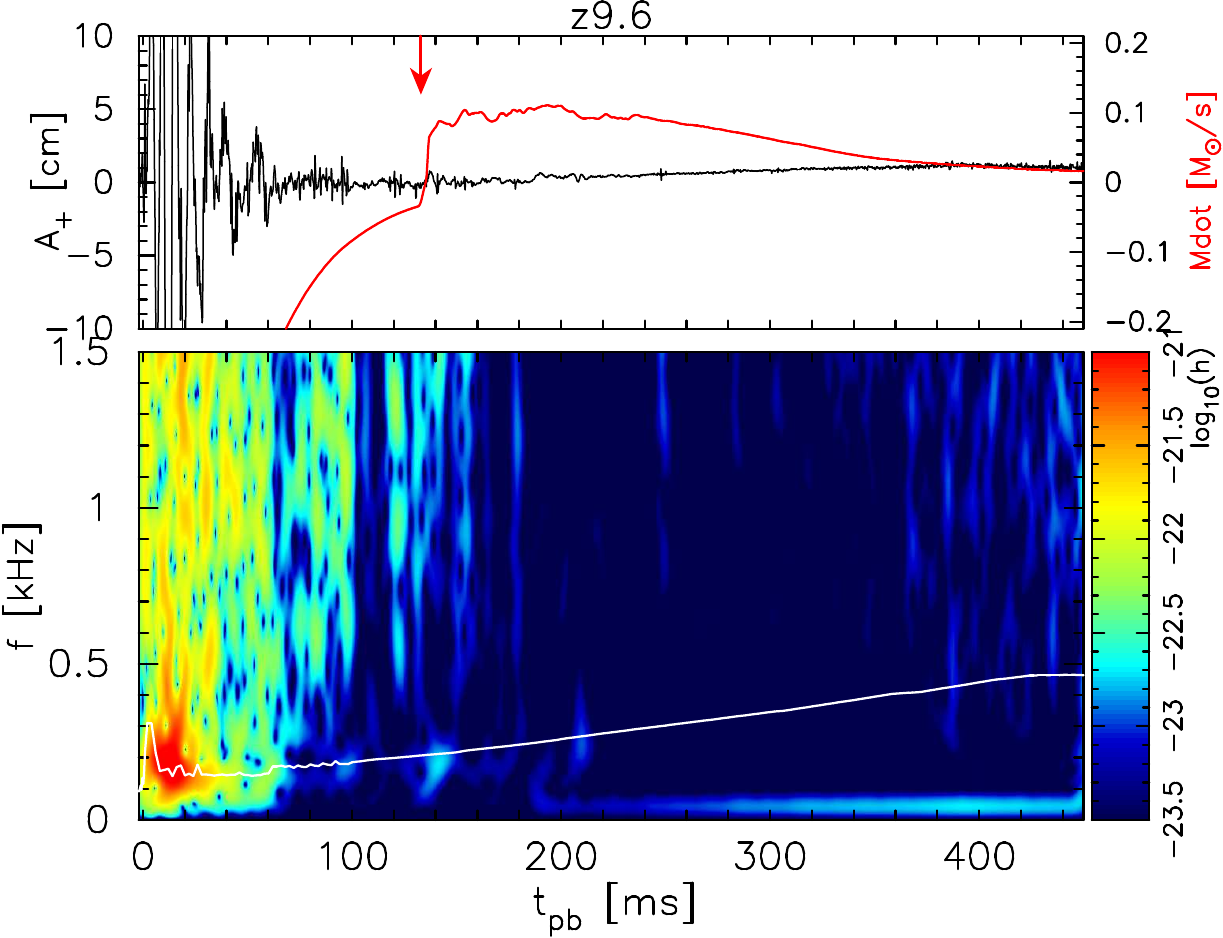}\label{fig:GW_1_a}}
\hspace{5mm}
\subfigure[]{\includegraphics[width=0.7\columnwidth,angle=0.]{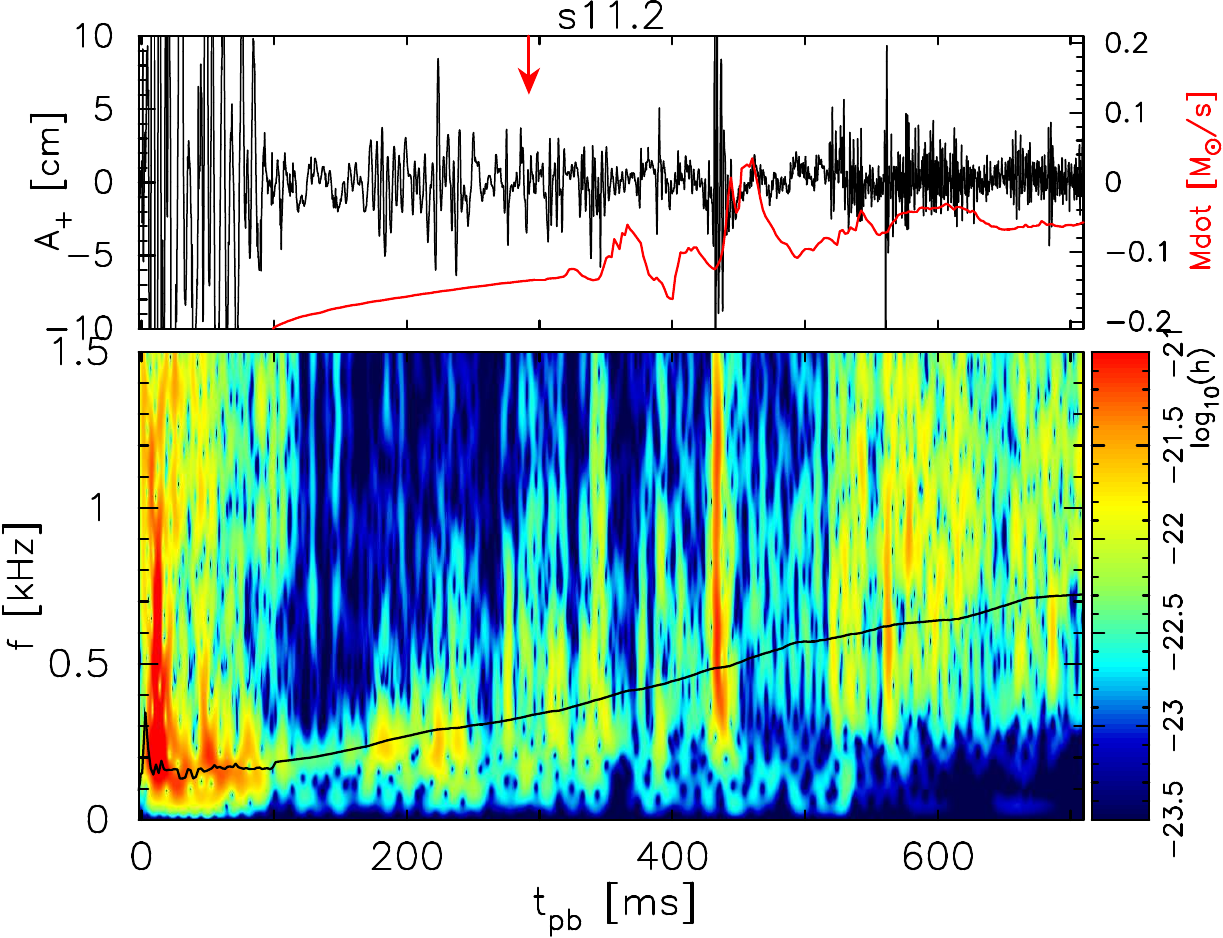}\label{fig:GW_1_b}}
\caption{GW waveform $A_+$ (black line in the top panels) and the corresponding spectrogram (colour coded in the bottom panels), mass accretion rate measured at $r=500$\,km (red line in the top panels), and the analytical expression for $f_{\rm peak}$ (black/white line in the bottom panels), for z9.6 (left) and s11.2 (right).
We show only the non-vanishing component of the axisymmetric profile $A_+\equiv Dh_+$ as seen along the equatorial plane, with $D$ and $h_+$ denoting the source distance and GW strain, respectively.
We also mark the time when the maximum shock radius reaches $500$\,km by a red arrow in the top panels.
Both figures are taken from \cite{KurodaT22}, but slightly modified.
\label{fig:GW_1}
}
\end{center}
\end{figure*}
Fig.~\ref{fig:GW_1} shows the GW waveform $A_+$ (black line, top panels) and its spectrogram (colour map, bottom panels), the mass accretion rate measured at $r=500$\,km (red line, top panels), and the analytical $f_{\rm peak}$ (black/white line, bottom panels; see \cite{BMuller13} for its definition), for z9.6 (left) and s11.2 (right).
Again, we show only the non-vanishing axisymmetric component $A_+\equiv Dh_+$, measured along the equatorial plane, which is perpendicular to the symmetry axis.
$D$ and $h_+$ are the source distance and GW strain, respectively, and $h_+$ is computed using a standard quadrupole formula.
The spectrogram is obtained using a short-time Fourier transform.
We also indicate, by a red arrow in the top panels, the time when the maximum shock radius reaches $500$\,km.
This helps us to tell whether the radius used to evaluate the mass accretion rate (red line) lies behind or ahead of the shock.

The main features of $A_+$ and its spectrogram are as follows.
During the initial convection phase ($t_{\rm pb}\lesssim50$–100~ms), relatively strong ($|A|\sim20$\,cm), low-frequency ($f\sim100$\,Hz) GWs are emitted.
In the later phase, z9.6 shows essentially no GW signal, except for an even lower-frequency component at $f\sim10$\,Hz for $t_{\rm pb}\gtrsim250$\,ms.
This low-frequency component is interpreted as a GW memory signal \citep{Christodoulou91}. This builds up due to the aspherical shock expansion, which in this case has a bipolar-like geometry (see the positive value $A_+\sim1$\,cm in Fig.~\ref{fig:GW_1_a} and Refs. \cite{Shibata06,martin08,simon10,shibagaki26,schnauck26} for a more extreme magnetorotational collapse model with jet formation). See also \cite{Vartanyan20,cecilia21,gill2024,choi24,colter24} for a memory GW generated by anisotropic neutrino emission, the detection of which requires a low-frequency GW detector $\lesssim 10$ Hz in the next generation including DECIGO (DECi-hertz Interferometer Gravitational wave Observatory) \cite{decigo}.
The quiescent phase in z9.6 most likely reflects the lack of an external driver strong enough to excite significant PNS oscillations.
As the mass accretion rate $\dot M$ (red line, top panels) clearly shows, $\dot M$ in z9.6 becomes positive at $t_{\rm pb}\gtrsim150$\,ms because of the explosion, or more precisely because of the pattern of outflow and inflow associated with the explosion.
We note that the extraction radius for the mass accretion rate (here $r=500$\,km) is chosen somewhat arbitrarily and has no special significance.
The important point is that $\dot M$ behind the shock is a more direct measure of how accretion excites PNS oscillations than $\dot M$ ahead of the shock.
Our discussion would not change if we instead used $r=300$ or 400\,km.
The positive $\dot M$ behind the shock in z9.6 indicates the absence of strong downflows.
Consequently, the PNS cannot sustain strong oscillations in the later phase, and the GWs remain weak, as seen in the spectrogram.
In contrast, in s11.2 the accretion rate $\dot M$ remains negative even in the post-explosion phase (see $r_{\rm shock}$ and $E_{\rm exp}$ in Fig.~\ref{fig:Evolution_Kuroda22}).
This is a consequence of global flow pattern behind the shock, comprised of mass outflow, which pushes the shock outward, and of mass accretion channel.
Because of the presence of latter, GW emission continues even after explosion.
Interestingly, such stochastic flow patterns sometimes produce a burst-like GW signal followed by a temporary quiet phase.
For example, at $t_{\rm pb}\sim430$\,ms we see a burst-like, relatively strong GW signal in s11.2.
In the following phase, $430\lesssim t_{\rm pb}\lesssim500$\,ms, the spectrogram shows somewhat weaker signals.
We attribute this behaviour to a sudden mass accretion episode onto the PNS, as first pointed out by \cite{BMuller13}.
First, a transient, strong accretion event occurs, which is supported by the fact that $\dot M$ becomes more negative before the GW burst (see the red line in s11.2 just before $t_{\rm pb}\sim430$\,ms).
Such behaviour is entirely due to stochastic flow patterns inside the PNS.
This accretion excites PNS oscillations and can, at the same time, trigger a small {\it explosion}, as part of the released gravitational energy is used to unbind matter.
As a result, the accretion rate in the later phase can decrease, and the net mass flux can even become positive, as in s11.2.
If such a significant decrease in $\dot M$ occurs, the perturbations acting on the PNS are reduced and the spectrogram shows weaker GW signals in the corresponding interval ($430\lesssim t_{\rm pb}\lesssim500$\,ms), similar to the quiescent phase in z9.6.
This example illustrates the tight and visible link between the explosion dynamics and the GW emission.

Next, we discuss the characteristic GW peak frequency.
The peak frequency is thought to depend mainly on the compactness of the PNS ($M/R$).
From the right panel of Fig.~\ref{fig:Evolution_Kuroda22}, $M/R$ shows essentially the same evolution for z9.6 and s11.2, at least up to $t_{\rm pb}\sim500$\,ms.
Using this $M/R$ evolution and the neutrino emission properties, we overlay $f_{\rm peak}$ \cite{BMuller13} on the GW spectrogram.
The $f_{\rm peak}$ curves for z9.6 and s11.2 have very similar shapes, consistent with their nearly identical $M/R$ values.
For example, $f_{\rm peak}$ at $t_{\rm pb}=400$\,ms is 424~Hz for z9.6 and 441~Hz for s11.2, close to the value in the more massive, BH-forming model z70 discussed in the next section.
For s11.2, the overall structure in the spectrogram follows the $f_{\rm peak}$ curve quite well, especially in the early phase ($t_{\rm pb}\lesssim300$\,ms), while at later times the spectrum becomes broader-band, making it harder to single out one peak frequency and compare it with $f_{\rm peak}$.
For z9.6, although there is no visible GW emission for $t_{\rm pb}\gtrsim100$\,ms, the early spectrogram is also consistent with the evolution of $f_{\rm peak}$.
Regarding kHz GW components, s11.2 shows burst-like GWs (e.g. at $t_{\rm pb}\sim430$\,ms), whose frequency content spans a wide range and extends above 1~kHz.
Another notable feature is that the overall GW power in the spectrogram shifts toward higher frequencies with time.
In particular, at $t_{\rm pb}\gtrsim520$\,ms, noisy GW signals reaching the kHz band are clearly visible.
Comparing these two SN models, we see that differences in $M/R$ evolution and mass accretion—and hence in the explosion dynamics—have a clear impact on the GW emission in standard CCSN scenarios.

\section{High frequency GWs from beyond-the-standard emission model: BH formation}
\label{sec:BH formation}
The previous section reviewed the standard CCSN GW emission mechanism, which is associated with canonical SN models, typically with explosions driven by neutrino heating.
In this section, we discuss how these standard GW signals change in non-standard cases.

\begin{figure*}[htbp]
\begin{center}
{\includegraphics[width=0.60\columnwidth,angle=0.]{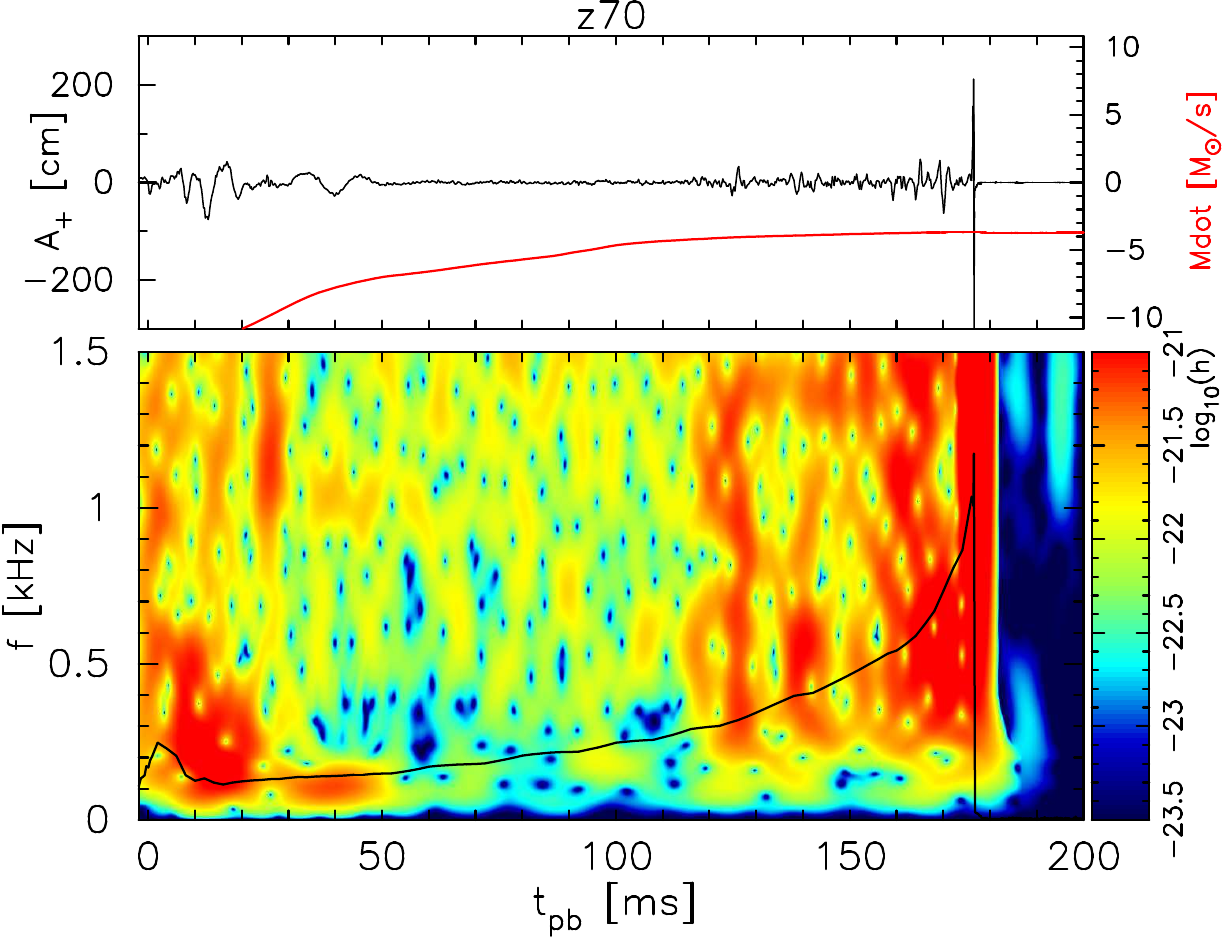}}
\hspace{5mm}\\
\caption{Same as Fig.~\ref{fig:GW_1}, but for our most massive progenitor model z70.
Taken from \cite{KurodaT22} with a slight modification.
\label{fig:GW_2}
}
\end{center}
\end{figure*}
We begin with the case of BH formation.
Detailed discussions of the BH formation dynamics and associated multi-messenger signals, in particular the neutrino signal, can be found in \cite{KurodaT23BH,KurodaT24}.
Fig.~\ref{fig:GW_2} is the same as Fig.~\ref{fig:GW_1}, but for our most massive progenitor, z70.
We note that the $y$-axis range in the top panel differs from that for the less massive progenitors in Fig.~\ref{fig:GW_1}.
The overall qualitative behaviour is similar to that of z9.6 and s11.2, except that the GW amplitudes $|A|$ and frequencies $f$ are generally larger.
During the initial convection phase ($t_{\rm pb}\lesssim50$\,ms), the GW amplitude $|A|$ reaches several tens of centimetres.
A quiescent phase then follows, also seen in this model.
The most striking difference from canonical CCSN models is the mass accretion rate $|\dot M|$.
Throughout the simulation, it always exceeds $\sim4\,M_\odot$\,s$^{-1}$, which is more than an order of magnitude larger than in the less massive models.
Such a high accretion rate rapidly compresses the PNS and shifts its typical oscillation frequencies to higher values.
This trend is clearly visible in the evolution of $f_{\rm peak}$ (black line) in the bottom panel.
Importantly, the $f_{\rm peak}$ curve agrees well with the main features in the GW spectrogram.
At $t_{\rm pb}\gtrsim120$\,ms the spectrogram is already dominated by broad-band emission from $f\sim200$\,Hz up to a few kHz, which likely reflects vigorous convective plumes penetrating into a rapidly contracting PNS and exciting several oscillation modes.
This GW emission stops abruptly at the time of BH formation,\footnote{In this study, we compute the GW strain only from the matter contribution using the conventional quadrupole formula \cite{Misner73,Moenchmeyer91}. The present waveform therefore contains no contribution from the BH itself. The ringdown signal from the newly formed BH is left for future work.} which occurs at $t_{\rm pb}=177$\,ms.

Compared with the less massive progenitors (z9.6 and s11.2), high-frequency GW emission thus appears much earlier.
For example, from Fig.~\ref{fig:GW_1_b}, s11.2 produces only low-frequency GWs at $f\sim300$\,Hz or less in the same time window $t_{\rm pb}\lesssim200$\,ms, apart from a brief high-frequency component associated with core bounce.
We can therefore, in principle, distinguish BH-forming models from standard ones by detecting high-frequency GW signals in such an early post-bounce phase, i.e. during the first few hundred milliseconds.
Moreover, an abrupt disappearance of the GW signal would provide further support for BH formation, as the central BH quickly engulfs the surrounding matter that had been producing the GWs.

\section{High frequency GWs from beyond-the-standard emission model: exotic compact star formation}
\label{sec:Exotic star formation}
\begin{figure*}[t!]
\centering
\includegraphics[width=0.7\columnwidth,angle=0]{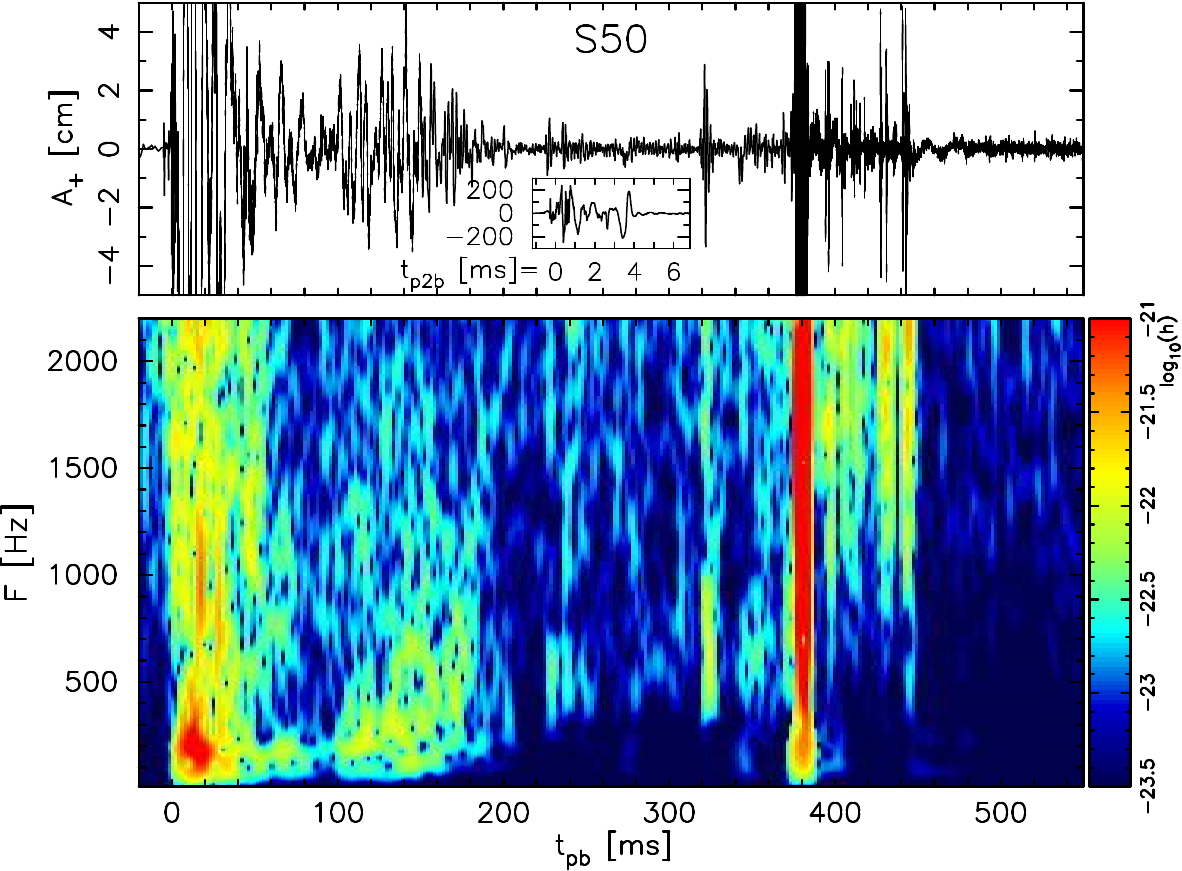}
\caption{Similar to Fig.~\ref{fig:GW_1}.
In the inner mini-panel of the top panel, we show a magnified view of the GW waveform around the second bounce time $t_{\rm p2b}$. Taken from \cite{KurodaT22}.
\label{fig:GW_3}}
\end{figure*}
So far, we have discussed how the collapse of a PNS into a BH can be a source of characteristic high-frequency GWs.
Recent advances in supernova theory have revealed several alternative models beyond the standard picture, which can also produce strong high-frequency GWs.
In this section, we review two such models that have been proposed only recently.
Although the number of studies is still small and the field is at an early stage, these models offer a way to link nuclear physics, alternatives to GR, and supernova theory through GW and neutrino observations.
They are therefore a particularly interesting and promising direction of research.
\subsection{Hybrid star formation}
\label{sec:Hybrid star formation}
In \cite{KurodaT22}, possible observable signatures of a first-order QCD phase transition in the context of CCSNe were studied in detail, where axisymmetric numerical-relativity simulations were conducted with multi-energy neutrino transport, using a hadron–quark hybrid EoS.
Four non-rotating progenitor models were taken, whose post-bounce evolution has already been outlined in Fig.~\ref{fig:Evolution_Kuroda22}.
Here, we focus on one of them, s50, which has a moderately high ZAMS mass of $50$\,$M_\odot$.
In this model, shock revival via the standard neutrino heating mechanism does not occur.
Instead, the PNS core enters the phase-transition region at $t_{\rm pb}\sim380$\,ms as its density rises due to contraction, and the core undergoes a second collapse.
Because of the sudden stiffening of the EoS when the core enters the pure quark-matter phase, a strong shock is launched (see the red solid line in the left panel of Fig.~\ref{fig:Evolution_Kuroda22}) and blows off the PNS envelope.
As a result, a quark core surrounded by hadronic matter is left behind, forming a hybrid star.
This captures the overall evolution of such moderately massive progenitors when a strong first-order phase transition is included.

We now discuss the GW signal associated with such a phase transition and how it differs from those from the canonical CCSN models.
Fig.~\ref{fig:GW_3} shows the GW waveform (upper panel) of model s50 and its spectrogram (lower panel), in a format similar to Fig.~\ref{fig:GW_1}.
In the inset of the upper panel, we show a magnified view of the waveform around the second bounce, where the $x$-axis measures the time relative to the second bounce, $t_{\rm p2b}$, in milliseconds.
The overall spectrogram evolution is essentially the same as in the other models (z9.6, s11.2, and z70) until the second collapse and phase transition at $t_{\rm pb}\sim380$\,ms.
When the phase transition occurs, the GW amplitude reaches $A_+\sim250$\,cm and the strong emission lasts for about 4\,ms after the second bounce.
Because of its burst-like nature, the emission spans a wide frequency range, from $\sim500$\,Hz up to beyond $2.5$\,kHz.
Ref.~\cite{KurodaT22} showed that these high-frequency components are comparable to the typical frequencies of convection behind the second-bounce shock.

\begin{figure}[htbp]
\begin{center}
\includegraphics[width=55mm,angle=-90.]{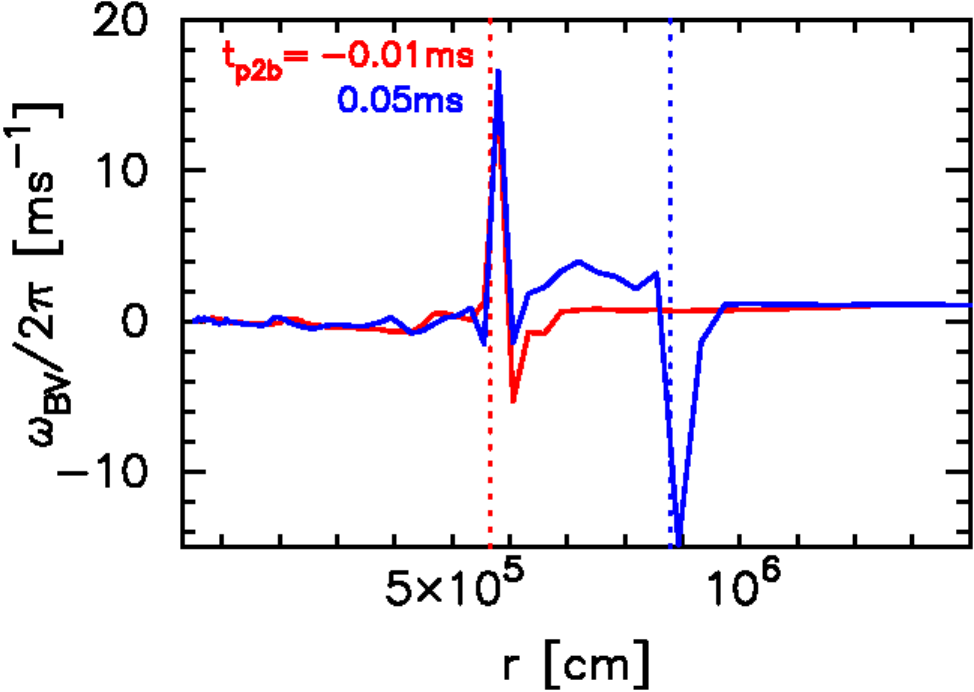}
\caption{Spherically averaged spatial profiles of the Brunt–V\"ais\"al\"a frequency $\omega_{\rm BV}$ for s50 around the second bounce.
We focus on the region above the quark-matter core and select two representative time slices at $t_{\rm p2b}=-0.01$\,ms (red line) and $t_{\rm p2b}=0.05$\,ms (blue line).
The vertical dashed lines indicate the average radii of the second-bounce shock.
Negative $\omega_{\rm BV}$ marks a convectively unstable region according to the Ledoux criterion.
Taken from \cite{KurodaT22}.
\label{fig:BVfrequency}}
\end{center}
\end{figure}

According to \cite{KurodaT22}, the second, strong shock launched by the quark-core bounce is produced at radii of a few kilometres from the centre (e.g. $r\sim6$\,km in model s50).
In a typical SN core structure ($r\lesssim10$\,km), the entropy profile $s(r)$ increases outward, i.e. $ds/dr\gtrsim0$, while the lepton fraction $Y_l(r)$ has a steep negative gradient, $dY_l/dr<0$.
The same behaviour is found in model s50 (see Fig.~5 in \cite{KurodaT22}).
The strong negative lepton gradient leads to convective instability.
To quantify this, we evaluate the Brunt–V\"ais\"al\"a frequency $\omega_{\rm BV}$ and plot its spherically averaged profile around the second bounce in Fig.~\ref{fig:BVfrequency}.
We follow \cite{BMuller13} and compute $\omega_{\rm BV}$ (see their Eq.~(14)) in the relativistic form
\begin{eqnarray}
\label{eq:BV1}
\omega_{\rm BV}^2=\frac{\alpha }{\rho h \phi^4}\frac{\partial \alpha}{\partial r}\left(\frac{\partial \rho(1+\epsilon)}{\partial r}-\frac{1}{c_s^2}\frac{\partial P}{\partial r}\right).
\end{eqnarray}
With this definition, regions with positive and negative $\omega_{\rm BV}^2$ correspond to convectively stable and unstable layers, respectively.
In Fig.~\ref{fig:BVfrequency}, we plot $\omega_{\rm BV}/2\pi$ by redefining $\omega_{\rm BV}$ as $\omega_{\rm BV}={\rm sgn}(\omega_{\rm BV}^2)\sqrt{|\omega_{\rm BV}^2|}$, so that it is a real quantity whose sign directly indicates convective stability (positive) or instability (negative).
We show two time slices around the second bounce, marked by the red ($t_{\rm p2b}=-0.01$\,ms) and blue ($t_{\rm p2b}=0.05$\,ms) curves.
During this short interval, the quark-core surface remains at $r\sim6\times10^5$\,cm, whereas the shock front moves from $r\sim6\times10^5$\,cm at $t_{\rm p2b}=-0.01$\,ms to $\sim9\times10^5$\,cm at $t_{\rm p2b}=0.05$\,ms.
We find that $\omega_{\rm BV}$ in front of the second-bounce shock is negative, indicating a convectively unstable region according to the Ledoux criterion, with $|\omega_{\rm BV}/2\pi|$ exceeding $\sim10$\,ms$^{-1}$.
Because of this large negative $\omega_{\rm BV}$, convection can grow quickly during the shock passage—within about a millisecond—and seeds further convective overturn.
Between the quark core and the second-bounce shock ($6\times10^5$\,cm$\lesssim r\lesssim9\times10^5$\,cm at $t_{\rm p2b}=0.05$\,ms), the characteristic angular frequencies (or “plume frequencies” \citep{Murphy09}) reach $\omega_{\rm BV}/2\pi\sim$ a few kHz.
These strong convective plumes behind the second shock are the main source of the high-frequency GWs seen in Fig.~\ref{fig:GW_3}, and the plume frequencies ($\omega_{\rm BV}/2\pi\sim$ a few kHz) can reasonably account for the GW frequency at the second bounce.
The maximum amplitude and duration of the strong emission agree well with those found by \cite{Zha20}.

\begin{figure}[htbp]
\begin{center}
\includegraphics[width=70mm,angle=0.]{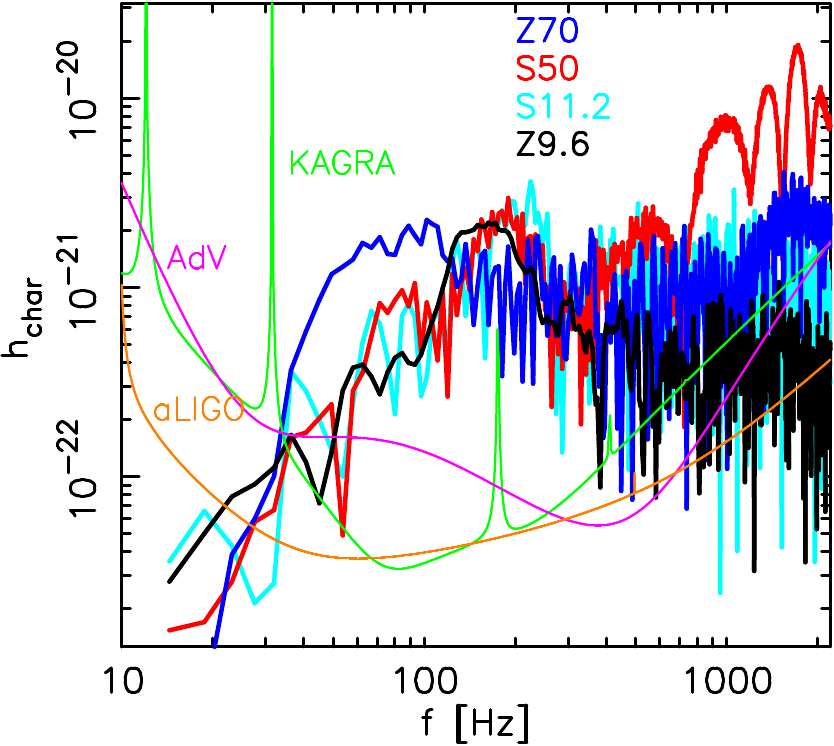}
\caption{Characteristic GW spectral amplitudes for the models discussed in Sections~\ref{sec:Standard GW emission mechanism}, \ref{sec:BH formation}, and \ref{sec:Hybrid star formation}, assuming a source distance of 10\,kpc.
The noise amplitudes of aLIGO (orange line), AdV (magenta line), and KAGRA (green line) are shown for comparison.
Taken from \cite{KurodaT22}.
\label{fig:hchar}}
\end{center}
\end{figure}
One goal of this section is to explore how a CCSN that undergoes a QCD phase transition can be distinguished from an ordinary CCSN by means of GW observations, if such a transition actually occurs.
To this end, Fig.~\ref{fig:hchar} shows the GW spectral amplitudes for all models discussed so far in Sections~\ref{sec:Standard GW emission mechanism}, \ref{sec:BH formation}, and \ref{sec:Hybrid star formation}, assuming a source distance of 10\,kpc, together with the sensitivity curves of Advanced LIGO (aLIGO), Advanced Virgo (AdV), and KAGRA \citep{abbott18det}.
The GW signal from model s50 (red line), which includes the phase transition, peaks in the kHz range, outside the most sensitive band of current ground-based GW detectors.
However, thanks to its large amplitude, it could still be detectable if the source is located within our Galaxy.
It is also clear that such strong kHz signals are absent in the standard SN models (z9.6 and s11.2).
Moreover, previous SN models that include the phase transition \cite{Sagert09,Fischer18,Zha20,KurodaT22,Andia23} show that an electron-type anti-neutrino burst accompanies the phase transition.
Therefore, the combined detection of strong kHz GWs (such as by the next-generation detectors (Cosmic Explorer \cite{CE} and Einstein Telescope \cite{ET}), see \cite{jade24} for several proposed high frequency detectors including NEMO \cite{nemo}) and a coincident neutrino burst can be a strong indication of such a second-collapse scenario, including the phase-transition-induced collapse of a PNS.

\subsection{Formation of compact star in beyond-GR theory}
\label{sec:Formation of compact star in beyond-GR theory}
Beyond the QCD–CCSN model, there are other, even less explored mechanisms for CCSN explosions.
Recently, we proposed a new explosion model within the framework of alternatives to GR \citep{KurodaT23STT}.
GR is the current standard theory of gravity and has passed many precision tests \citep{GR_Test_Will14,GR_Test_Psaltis08,GR_Test_berti15}.
However, the need to explain cosmic inflation and the fact that most of the energy content of the universe appears as dark energy and dark matter \citep{Perlmutter99,Komatsu7yrs,Abbott19,PlanckCollaboration20} motivate the study of alternatives to GR.
ST (scalar tensor) theories of gravity—one of the simplest and best motivated alternatives in cosmology and astrophysics—introduce an additional scalar field that allows sizeable deviations from GR in the strong-field regime.
A key phenomenon in ST theories is spontaneous scalarization (SS) \citep{Damour93,Ramazanoglu16}, in which non-linear coupling between the scalar and gravitational fields leads to an exponential growth of the scalar field.
If this were to occur inside an NS, its structure could change significantly.
Applying ST theories to core collapse, Refs.~\cite{Novak00,Gerosa16,Rosca-Mead20} showed that SS can also occur in PNSs, leading to large structural changes, strong shock formation, and the emission of potentially detectable scalar-type GWs.

\begin{figure}[t!]
\begin{center}
\includegraphics[angle=0.,width=0.6\columnwidth]{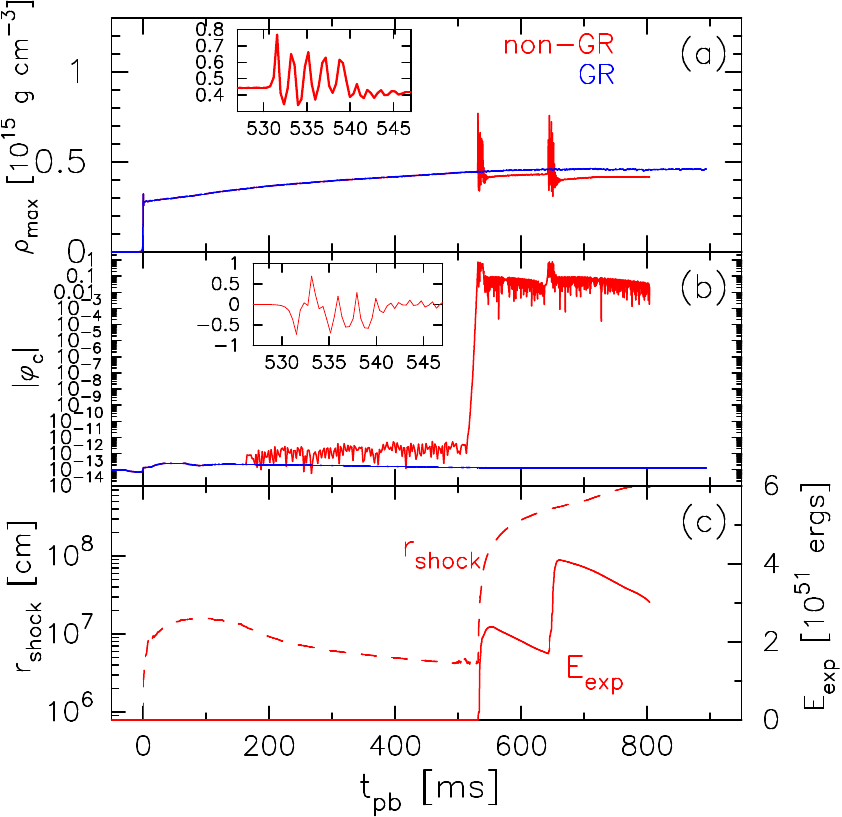} 
\caption{Overall evolution of all models. Panel (a): maximum rest-mass density $\rho_{\rm max}$; (b): PNS rest mass $M_{\rm PNS}$ and central lapse function $\alpha_{\rm c}$; (c): central scalar field $\varphi_{\rm c}$ and 2-norm of the Hamiltonian constraint $||\mathcal{H}||_2\times10^{-4}$; (d): averaged shock radius $R_{\rm s}$ (dashed) and diagnostic explosion energy $E_{\rm exp}$ (solid).
The colours identify the different models listed in panel (a).
Original figure from \cite{KurodaT23STT}.
\label{fig:Overall_Kuroda23}}
\end{center}
\end{figure}
In this context, Ref.\cite{KurodaT23STT} reported results from   multi-dimensional (2D) CCSN simulations in a {\it massive} ST (MST) theory for the first time, using a realistic EoS and multi-energy neutrino radiation.
The overall evolution is summarized in Fig.~\ref{fig:Overall_Kuroda23}.
It shows: (a) the maximum rest-mass density $\rho_{\rm max}$; (b) the PNS rest mass $M_{\rm PNS}$ and central lapse function $\alpha_{\rm c}$; (c) the central scalar field $\varphi_{\rm c}$ and the 2-norm of the Hamiltonian constraint $||\mathcal{H}||_2\times10^{-4}$; and (d) the averaged shock radius $R_{\rm s}$ and diagnostic explosion energy $E_{\rm exp}$.
Full model details are given in \cite{KurodaT23STT}; here we focus on two models, labelled ``beyond-GR'' (model B20M11 in the original paper) and ``GR'' (model B10M14).
The former shows a significant deviation from GR, while the latter stays essentially in the GR limit throughout the simulation even though the simulation itself was carried out in the ST theory.
Insets in panels (a) and (b) present magnified views around the first SS event.

A key feature for high-frequency GW emission is the second and even third collapse of the PNS, which occur at $t_{\rm pb}\sim530$ and 650\,ms, respectively.
From panel (c), we see that the scalar field $\varphi$ rapidly deviates from its far-zone value $|\varphi|\ll1$ by many orders of magnitude, signalling a departure from GR.
This marks the onset of SS \cite{Damour93,Shibata_JFBD_14}.
The deviation from GR strongly changes the internal structure of the PNS and produces multiple collapses and bounces, each of which drives an explosion and ejects part of the PNS envelope, as shown in panel (d).

\begin{figure}[t!]
\begin{center}
\includegraphics[angle=-0.,width=1.\columnwidth]{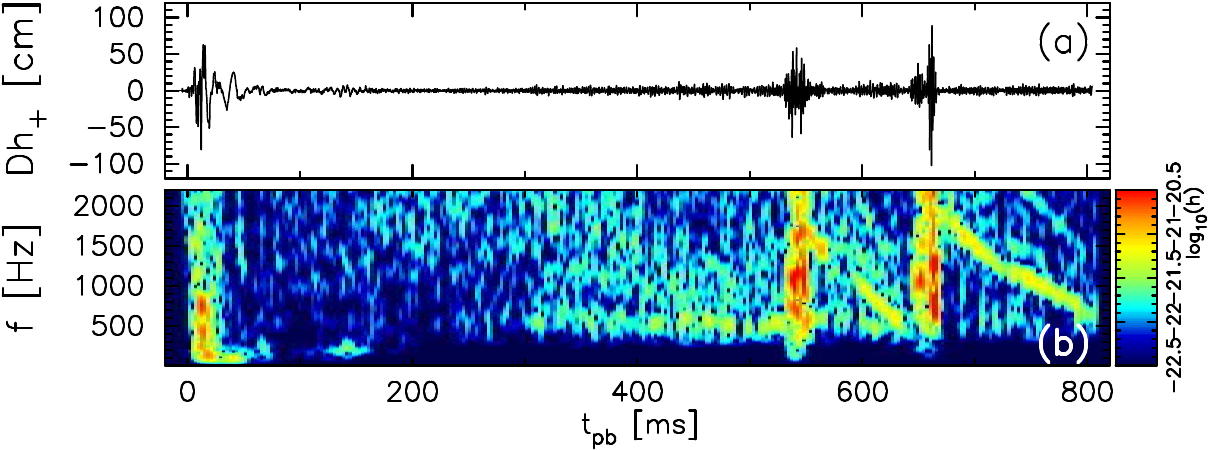}  
\caption{Panel (a): matter-origin GWs $Dh_+$; (b): spectrogram $\tilde h(f,t)$ of $h_+$. We assume a source distance of $D=10$\,kpc.
Modified from the original version in \cite{KurodaT23STT}.
Taken from \cite{KurodaT23STT}, but slightly modified.
\label{fig:GW_Kuroda23}}
\end{center}
\end{figure}

The GW signal from these multiple SS episodes is presented in Fig.~\ref{fig:GW_Kuroda23}.
Panel (a) shows the GW waveform, and panel (b) shows the spectrogram, in a format similar to Fig.~\ref{fig:GW_1}, but now for the ``beyond-GR'' model.
Owing to essentially the same mechanism as in the QCD model in Fig.~\ref{fig:GW_3}, the second and third collapses and bounces produce strong GWs with frequencies extending beyond $\sim2$\,kHz.
We again interpret these high-frequency components as being driven by vigorous convection behind the second and third shocks, analogous to the QCD case.
At present, there are very few simulations of either the QCD-transition model or the beyond-GR model.
It is therefore not yet clear whether future GW and neutrino observations will be able to distinguish these two types of events, both involving the gravitational collapse of a PNS.
However, if more simulations across a wider parameter space become available, our understanding will improve significantly.
In addition, if quantities such as the density at which the QCD phase transition starts, the density gap of the mixed phase, or as-yet-unknown parameters in ST theories (for example the scalar mass) are better constrained, it may become possible to tell these scenarios apart using multi-messenger observations \citep[see also][for a brief discussion of possible differences between the QCD and beyond-GR models]{KurodaT23STT}.

\section{Summary} 
\label{sec:Conclusion}
We have systematically reviewed the high-frequency GW emission mechanisms from CCSNe, focusing both on standard-CCSN-model expectations and a range of beyond-standard SN scenarios.
State-of-the-art multidimensional simulations \citep{Nakamura22,KurodaT22} demonstrate that standard GWs primarily originate from the excitation of oscillation modes of the PNS, notably the $g$- and $f$-modes.
The characteristic GW frequency is closely tied to the evolving compactness parameter $M/R$ of the PNS, and the amplitude is regulated by the mass accretion history, reflecting the underlying progenitor structure and explosion dynamics \citep{BMuller13,Murphy09,Powell21}.
Thus, observations of the temporal evolution of GW frequencies and amplitudes offer a direct probe into the extreme physical state, the explosion mechanism, and the progenitor’s properties.

Beyond-standard cases, such as PNS collapse to a BH or the formation of exotic compact objects, amplify and diversify the GW signals.
Our review highlights that BH formation leads to an early and rapid emergence of high-frequency GWs (at $\sim$\,several kHz), followed by the sudden subsidence of emission as mass accretion overwhelms the PNS and a BH forms \citep{KurodaT23BH}.
We further discussed the impact of first-order QCD phase transitions within the PNS, which induce multiple collapses and bounces, significantly enhancing both GW amplitudes and their characteristic frequencies up to several kHz \citep{KurodaT22,Zha20}.
Importantly, such phase transitions would be accompanied by other observable messengers, such as a distinct anti-neutrino burst, providing clear signatures to distinguish these events from the ordinary CCSN scenario.

Going even further, models invoking ST theories of gravity predict SS events and associated multiple collapses and bounces of the PNS, again producing strong, broadband, high-frequency GW emission \citep{KurodaT23STT,Damour93}.
These signatures bear remarkable similarities to QCD-induced phenomena, emphasizing that future GW and multi-messenger observations will be critical for disentangling different kinds of new physics in the extreme-density regime.
Systematic comparison of GW spectrograms, temporal evolution of frequencies, and correlated neutrino emissions across models will be essential for discriminating among different explosion mechanisms and constraining novel physical parameters, including the nuclear EoS and possible deviations from GR.

In summary, high-frequency GWs from CCSNe serve as invaluable, direct probes into the inner dynamics of stellar explosions, revealing the interplay between progenitor structure, the EoS, and the possible emergence of new phases of matter or new gravitational phenomena.
As GW detectors improve in sensitivity and frequency range, particularly in the kHz regime (e.g., \cite{ET,CE,nemo}), the prospects for robustly distinguishing between standard and exotic core-collapse scenarios grow increasingly promising.
The synergetic analysis of GW and neutrino signals, informed by continued advances in numerical relativity and nuclear astrophysics, will open up unprecedented windows into extreme matter and gravity \citep{BMuller13,Mezzacappa23,Lella26}.

\ack{We are grateful to Tomoya Takiwaki, Ko Nakamura, Shota Shibagaki, Jin Matsumoto,  Kanji Mori, Tobias Fischer, and Masaru Shibata for stimulating discussions, which has enabled us to write this review article. KK expresses thanks to Shunsaku Horiuchi, Tomoya Kinugawa, Yohei Masada, Tsurugi Takata, and Adam Burrows for helpful exchanges. Numerical computations were carried out on Cray XC50 at the Center for Computational Astrophysics, National Astronomical Observatory of Japan and also on Sakura and Raven at Max Planck Computing and Data Facility.}

\funding{This work
was supported by JSPS KAKENHI Grant Numbers (JP24K00631, JP26K07093), funding from Fukuoka University (Grant No.GR2606) and, MEXT as “Program for Promoting researches on the Supercomputer Fugaku”
(Structure and Evolution of the Universe Unraveled by Fusion of Simulation
and AI; Grant Number JPMXP1020230406), and JICFuS.}


\data{All data that support the findings of this study can be provided upon request (and any supplementary
files).}


\bibliographystyle{unsrt}
\bibliography{main}

@ARTICLE{Bruenn04,
       author = {{Bruenn}, S.~W. and {Raley}, E.~A. and {Mezzacappa}, A.},
        title = "{Fluid Stability Below the Neutrinospheres of Supernova Progenitors and the Dominant Role of Lepto-Entropy Fingers}",
      journal = {arXiv e-prints},
         year = 2004,
        month = apr,
          eid = {astro-ph/0404099},
        pages = {astro-ph/0404099},
          doi = {10.48550/arXiv.astro-ph/0404099},
archivePrefix = {arXiv},
       eprint = {astro-ph/0404099},
 primaryClass = {astro-ph},
       adsurl = {https://ui.adsabs.harvard.edu/abs/2004astro.ph..4099B}
}

@ARTICLE{urushibata,
       author = {{Urushibata}, Takaki and {Takahashi}, Koh and {Umeda}, Hideyuki and {Yoshida}, Takashi},
        title = "{A progenitor model of SN 1987A based on the slow-merger scenario}",
      journal = {\mnras},
         year = 2018,
        month = jan,
       volume = {473},
       number = {1},
        pages = {L101-L105},
          doi = {10.1093/mnrasl/slx166},
archivePrefix = {arXiv},
       eprint = {1705.04084},
 primaryClass = {astro-ph.SR},
       adsurl = {https://ui.adsabs.harvard.edu/abs/2018MNRAS.473L.101U}
}

@article{takahara,
  title={Can phase transitions of superdense matter strengthen supernova explosions},
  author={Mariko Takahara and Katsuhiko Sato},
  journal={Physics Letters B},
  year={1985},
  volume={156},
  pages={17-21},
  url={https://api.semanticscholar.org/CorpusID:120922060}
}

@ARTICLE{gentile,
       author = {{Gentile}, N.~A. and {Aufderheide}, M.~B. and {Mathews}, G.~J. and {Swesty}, F.~D. and {Fuller}, G.~M.},
        title = "{The QCD Phase Transition and Supernova Core Collapse}",
      journal = {\apj},
         year = 1993,
        month = sep,
       volume = {414},
        pages = {701},
          doi = {10.1086/173116},
       adsurl = {https://ui.adsabs.harvard.edu/abs/1993ApJ...414..701G}
}

@ARTICLE{jim15,
       author = {{Fuller}, Jim and {Klion}, Hannah and {Abdikamalov}, Ernazar and {Ott}, Christian D.},
        title = "{Supernova seismology: gravitational wave signatures of rapidly rotating core collapse}",
      journal = {\mnras},
         year = 2015,
        month = jun,
       volume = {450},
       number = {1},
        pages = {414-427},
          doi = {10.1093/mnras/stv698},
archivePrefix = {arXiv},
       eprint = {1501.06951},
 primaryClass = {astro-ph.HE},
       adsurl = {https://ui.adsabs.harvard.edu/abs/2015MNRAS.450..414F}
}

@ARTICLE{bruel23,
       author = {{Bruel}, Tristan and {Bizouard}, Marie-Anne and {Obergaulinger}, Martin and {Maturana-Russel}, Patricio and {Torres-Forn{\'e}}, Alejandro and {Cerd{\'a}-Dur{\'a}n}, Pablo and {Christensen}, Nelson and {Font}, Jos{\'e} A. and {Meyer}, Renate},
        title = "{Inference of protoneutron star properties in core-collapse supernovae from a gravitational-wave detector network}",
      journal = {\prd},
         year = 2023,
        month = apr,
       volume = {107},
       number = {8},
          eid = {083029},
        pages = {083029},
          doi = {10.1103/PhysRevD.107.083029},
archivePrefix = {arXiv},
       eprint = {2301.10019},
 primaryClass = {astro-ph.HE},
       adsurl = {https://ui.adsabs.harvard.edu/abs/2023PhRvD.107h3029B}
}

@ARTICLE{marie_anne21,
       author = {{Bizouard}, Marie-Anne and {Maturana-Russel}, Patricio and {Torres-Forn{\'e}}, Alejandro and {Obergaulinger}, Martin and {Cerd{\'a}-Dur{\'a}n}, Pablo and {Christensen}, Nelson and {Font}, Jos{\'e} A. and {Meyer}, Renate},
        title = "{Inference of protoneutron star properties from gravitational-wave data in core-collapse supernovae}",
      journal = {\prd},
         year = 2021,
        month = mar,
       volume = {103},
       number = {6},
          eid = {063006},
        pages = {063006},
          doi = {10.1103/PhysRevD.103.063006},
archivePrefix = {arXiv},
       eprint = {2012.00846},
 primaryClass = {gr-qc},
       adsurl = {https://ui.adsabs.harvard.edu/abs/2021PhRvD.103f3006B}
}

@ARTICLE{colter22,
       author = {{Richardson}, Colter J. and {Zanolin}, Michele and {Andresen}, Haakon and {Szczepa{\'n}czyk}, Marek J. and {Gill}, Kiranjyot and {Wongwathanarat}, Annop},
        title = "{Modeling core-collapse supernovae gravitational-wave memory in laser interferometric data}",
      journal = {\prd},
         year = 2022,
        month = may,
       volume = {105},
       number = {10},
          eid = {103008},
        pages = {103008},
          doi = {10.1103/PhysRevD.105.103008},
archivePrefix = {arXiv},
       eprint = {2109.01582},
 primaryClass = {astro-ph.HE},
       adsurl = {https://ui.adsabs.harvard.edu/abs/2022PhRvD.105j3008R}
}

@ARTICLE{decigo,
       author = {{Kawamura}, Seiji and {Ando}, Masaki and {Seto}, Naoki and {Sato}, Shuichi and {Musha}, Mitsuru and {Kawano}, Isao and {Yokoyama}, Jun'ichi and {Tanaka}, Takahiro and {Ioka}, Kunihito and {Akutsu}, Tomotada and {Takashima}, Takeshi and {Agatsuma}, Kazuhiro and {Araya}, Akito and {Aritomi}, Naoki and {Asada}, Hideki and {Chiba}, Takeshi and {Eguchi}, Satoshi and {Enoki}, Motohiro and {Fujimoto}, Masa-Katsu and {Fujita}, Ryuichi and {Futamase}, Toshifumi and {Harada}, Tomohiro and {Hayama}, Kazuhiro and {Himemoto}, Yoshiaki and {Hiramatsu}, Takashi and {Hong}, Feng-Lei and {Hosokawa}, Mizuhiko and {Ichiki}, Kiyotomo and {Ikari}, Satoshi and {Ishihara}, Hideki and {Ishikawa}, Tomohiro and {Itoh}, Yousuke and {Ito}, Takahiro and {Iwaguchi}, Shoki and {Izumi}, Kiwamu and {Kanda}, Nobuyuki and {Kanemura}, Shinya and {Kawazoe}, Fumiko and {Kobayashi}, Shiho and {Kohri}, Kazunori and {Kojima}, Yasufumi and {Kokeyama}, Keiko and {Kotake}, Kei and {Kuroyanagi}, Sachiko and {Maeda}, Kei-ichi and {Matsushita}, Shuhei and {Michimura}, Yuta and {Morimoto}, Taigen and {Mukohyama}, Shinji and {Nagano}, Koji and {Nagano}, Shigeo and {Naito}, Takeo and {Nakamura}, Kouji and {Nakamura}, Takashi and {Nakano}, Hiroyuki and {Nakao}, Kenichi and {Nakasuka}, Shinichi and {Nakayama}, Yoshinori and {Nakazawa}, Kazuhiro and {Nishizawa}, Atsushi and {Ohkawa}, Masashi and {Oohara}, Kenichi and {Sago}, Norichika and {Saijo}, Motoyuki and {Sakagami}, Masaaki and {Sakai}, Shin-ichiro and {Sato}, Takashi and {Shibata}, Masaru and {Shinkai}, Hisaaki and {Shoda}, Ayaka and {Somiya}, Kentaro and {Sotani}, Hajime and {Takahashi}, Ryutaro and {Takahashi}, Hirotaka and {Akiteru}, Takamori and {Taniguchi}, Keisuke and {Taruya}, Atsushi and {Tsubono}, Kimio and {Tsujikawa}, Shinji and {Ueda}, Akitoshi and {Ueda}, Ken-ichi and {Watanabe}, Izumi and {Yagi}, Kent and {Yamada}, Rika and {Yokoyama}, Shuichiro and {Yoo}, Chul-Moon and {Zhu}, Zong-Hong},
        title = "{Current status of space gravitational wave antenna DECIGO and B-DECIGO}",
      journal = {Progress of Theoretical and Experimental Physics},
         year = 2021,
        month = may,
       volume = {2021},
       number = {5},
          eid = {05A105},
        pages = {05A105},
          doi = {10.1093/ptep/ptab019},
archivePrefix = {arXiv},
       eprint = {2006.13545},
 primaryClass = {gr-qc},
       adsurl = {https://ui.adsabs.harvard.edu/abs/2021PTEP.2021eA105K}
}

@ARTICLE{colter24,
       author = {{Richardson}, Colter J. and {Andresen}, Haakon and {Mezzacappa}, Anthony and {Zanolin}, Michele and {Benjamin}, Michael G. and {Marronetti}, Pedro and {Lentz}, Eric J. and {Szczepa{\'n}czyk}, Marek J.},
        title = "{Detecting Gravitational Wave Memory in the Next Galactic Core-Collapse Supernova}",
      journal = {\prl},
         year = 2024,
        month = dec,
       volume = {133},
       number = {23},
          eid = {231401},
        pages = {231401},
          doi = {10.1103/PhysRevLett.133.231401},
archivePrefix = {arXiv},
       eprint = {2404.02131},
 primaryClass = {astro-ph.HE},
       adsurl = {https://ui.adsabs.harvard.edu/abs/2024PhRvL.133w1401R}
}

@ARTICLE{tony2024,
       author = {{Mezzacappa}, Anthony and {Zanolin}, Michele},
        title = "{Gravitational Waves from Neutrino-Driven Core Collapse Supernovae: Predictions, Detection, and Parameter Estimation}",
      journal = {arXiv e-prints},
         year = 2024,
        month = jan,
          eid = {arXiv:2401.11635},
        pages = {arXiv:2401.11635},
          doi = {10.48550/arXiv.2401.11635},
archivePrefix = {arXiv},
       eprint = {2401.11635},
 primaryClass = {astro-ph.HE},
       adsurl = {https://ui.adsabs.harvard.edu/abs/2024arXiv240111635M}
}

@ARTICLE{gill2024,
       author = {{Gill}, Kiranjyot},
        title = "{Milli-to-Deci-Hertz Detection Prospects for Gravitational Waves from Core-Collapse Supernovae}",
      journal = {arXiv e-prints},
         year = 2024,
        month = may,
          eid = {arXiv:2405.13211},
        pages = {arXiv:2405.13211},
          doi = {10.48550/arXiv.2405.13211},
archivePrefix = {arXiv},
       eprint = {2405.13211},
 primaryClass = {astro-ph.HE},
       adsurl = {https://ui.adsabs.harvard.edu/abs/2024arXiv240513211G}
}

@ARTICLE{jakobus_2023,
       author = {{Jakobus}, Pia and {M{\"u}ller}, Bernhard and {Heger}, Alexander and {Zha}, Shuai and {Powell}, Jade and {Motornenko}, Anton and {Steinheimer}, Jan and {St{\"o}cker}, Horst},
        title = "{Gravitational Waves from a Core g Mode in Supernovae as Probes of the High-Density Equation of State}",
      journal = {\prl},
         year = 2023,
        month = nov,
       volume = {131},
       number = {19},
          eid = {191201},
        pages = {191201},
          doi = {10.1103/PhysRevLett.131.191201},
archivePrefix = {arXiv},
       eprint = {2301.06515},
 primaryClass = {astro-ph.HE},
       adsurl = {https://ui.adsabs.harvard.edu/abs/2023PhRvL.131s1201J}
}

@ARTICLE{yamada_review,
       author = {{Yamada}, Shoichi and {Nagakura}, Hiroki and {Akaho}, Ryuichiro and {Harada}, Akira and {Furusawa}, Shun and {Iwakami}, Wakana and {Okawa}, Hirotada and {Matsufuru}, Hideo and {Sumiyoshi}, Kohsuke},
        title = "{Physical mechanism of core-collapse supernovae that neutrinos drive}",
      journal = {Proceedings of the Japan Academy, Series B},
         year = 2024,
        month = mar,
       volume = {100},
       number = {3},
        pages = {190-233},
          doi = {10.2183/pjab.100.015},
       adsurl = {https://ui.adsabs.harvard.edu/abs/2024PJAB..100..190Y}
}

@ARTICLE{Shibata06,
       author = {{Shibata}, Masaru and {Liu}, Yuk Tung and {Shapiro}, Stuart L. and {Stephens}, Branson C.},
        title = "{Magnetorotational collapse of massive stellar cores to neutron stars: Simulations in full general relativity}",
      journal = {\prd},
         year = 2006,
        month = nov,
       volume = {74},
       number = {10},
          eid = {104026},
        pages = {104026},
          doi = {10.1103/PhysRevD.74.104026},
archivePrefix = {arXiv},
       eprint = {astro-ph/0610840},
 primaryClass = {astro-ph},
       adsurl = {https://ui.adsabs.harvard.edu/abs/2006PhRvD..74j4026S}
}

@article{Christodoulou91,
  title = {Nonlinear nature of gravitation and gravitational-wave experiments},
  author = {Christodoulou, Demetrios},
  journal = {Phys. Rev. Lett.},
  volume = {67},
  issue = {12},
  pages = {1486--1489},
  numpages = {0},
  year = {1991},
  month = {Sep},
  publisher = {American Physical Society},
  doi = {10.1103/PhysRevLett.67.1486},
  url = {https://link.aps.org/doi/10.1103/PhysRevLett.67.1486}
}

@ARTICLE{tony_review,
       author = {{Mezzacappa}, Anthony and {Endeve}, Eirik and {Messer}, O.~E. Bronson and {Bruenn}, Stephen W.},
        title = "{Physical, numerical, and computational challenges of modeling neutrino transport in core-collapse supernovae}",
      journal = {Living Reviews in Computational Astrophysics},
         year = 2020,
        month = dec,
       volume = {6},
       number = {1},
          eid = {4},
        pages = {4},
          doi = {10.1007/s41115-020-00010-8},
archivePrefix = {arXiv},
       eprint = {2010.09013},
 primaryClass = {astro-ph.HE},
       adsurl = {https://ui.adsabs.harvard.edu/abs/2020LRCA....6....4M}
}

@ARTICLE{Torres-Forne18,
       author = {{Torres-Forn{\'e}}, Alejandro and {Cerd{\'a}-Dur{\'a}n}, Pablo and {Passamonti}, Andrea and {Font}, Jos{\'e} A.},
        title = "{Towards asteroseismology of core-collapse supernovae with gravitational-wave observations - I. Cowling approximation}",
      journal = {\mnras},
         year = 2018,
        month = mar,
       volume = {474},
       number = {4},
        pages = {5272-5286},
          doi = {10.1093/mnras/stx3067},
archivePrefix = {arXiv},
       eprint = {1708.01920},
 primaryClass = {astro-ph.SR},
       adsurl = {https://ui.adsabs.harvard.edu/abs/2018MNRAS.474.5272T}
}

@ARTICLE{CE,
       author = {{Evans}, Matthew and {Adhikari}, Rana X and {Afle}, Chaitanya and {Ballmer}, Stefan W. and {Biscoveanu}, Sylvia and {Borhanian}, Ssohrab and {Brown}, Duncan A. and {Chen}, Yanbei and {Eisenstein}, Robert and {Gruson}, Alexandra and {Gupta}, Anuradha and {Hall}, Evan D. and {Huxford}, Rachael and {Kamai}, Brittany and {Kashyap}, Rahul and {Kissel}, Jeff S. and {Kuns}, Kevin and {Landry}, Philippe and {Lenon}, Amber and {Lovelace}, Geoffrey and {McCuller}, Lee and {Ng}, Ken K.~Y. and {Nitz}, Alexander H. and {Read}, Jocelyn and {Sathyaprakash}, B.~S. and {Shoemaker}, David H. and {Slagmolen}, Bram J.~J. and {Smith}, Joshua R. and {Srivastava}, Varun and {Sun}, Ling and {Vitale}, Salvatore and {Weiss}, Rainer},
        title = "{A Horizon Study for Cosmic Explorer: Science, Observatories, and Community}",
      journal = {arXiv e-prints},
         year = 2021,
        month = sep,
          eid = {arXiv:2109.09882},
        pages = {arXiv:2109.09882},
          doi = {10.48550/arXiv.2109.09882},
archivePrefix = {arXiv},
       eprint = {2109.09882},
 primaryClass = {astro-ph.IM},
       adsurl = {https://ui.adsabs.harvard.edu/abs/2021arXiv210909882E}
}

@ARTICLE{ET,
       author = {{Hild}, S. and {Abernathy}, M. and {Acernese}, F. and {Amaro-Seoane}, P. and {Andersson}, N. and {Arun}, K. and {Barone}, F. and {Barr}, B. and {Barsuglia}, M. and {Beker}, M. and {Beveridge}, N. and {Birindelli}, S. and {Bose}, S. and {Bosi}, L. and {Braccini}, S. and {Bradaschia}, C. and {Bulik}, T. and {Calloni}, E. and {Cella}, G. and {Chassande Mottin}, E. and {Chelkowski}, S. and {Chincarini}, A. and {Clark}, J. and {Coccia}, E. and {Colacino}, C. and {Colas}, J. and {Cumming}, A. and {Cunningham}, L. and {Cuoco}, E. and {Danilishin}, S. and {Danzmann}, K. and {De Salvo}, R. and {Dent}, T. and {De Rosa}, R. and {Di Fiore}, L. and {Di Virgilio}, A. and {Doets}, M. and {Fafone}, V. and {Falferi}, P. and {Flaminio}, R. and {Franc}, J. and {Frasconi}, F. and {Freise}, A. and {Friedrich}, D. and {Fulda}, P. and {Gair}, J. and {Gemme}, G. and {Genin}, E. and {Gennai}, A. and {Giazotto}, A. and {Glampedakis}, K. and {Gr{\"a}f}, C. and {Granata}, M. and {Grote}, H. and {Guidi}, G. and {Gurkovsky}, A. and {Hammond}, G. and {Hannam}, M. and {Harms}, J. and {Heinert}, D. and {Hendry}, M. and {Heng}, I. and {Hennes}, E. and {Hough}, J. and {Husa}, S. and {Huttner}, S. and {Jones}, G. and {Khalili}, F. and {Kokeyama}, K. and {Kokkotas}, K. and {Krishnan}, B. and {Li}, T.~G.~F. and {Lorenzini}, M. and {L{\"u}ck}, H. and {Majorana}, E. and {Mandel}, I. and {Mandic}, V. and {Mantovani}, M. and {Martin}, I. and {Michel}, C. and {Minenkov}, Y. and {Morgado}, N. and {Mosca}, S. and {Mours}, B. and {M\"uller-Ebhardt}, H. and {Murray}, P. and {Nawrodt}, R. and {Nelson}, J. and {Oshaughnessy}, R. and {Ott}, C.~D. and {Palomba}, C. and {Paoli}, A. and {Parguez}, G. and {Pasqualetti}, A. and {Passaquieti}, R. and {Passuello}, D. and {Pinard}, L. and {Plastino}, W. and {Poggiani}, R. and {Popolizio}, P. and {Prato}, M. and {Punturo}, M. and {Puppo}, P. and {Rabeling}, D. and {Rapagnani}, P. and {Read}, J. and {Regimbau}, T. and {Rehbein}, H. and {Reid}, S. and {Ricci}, F. and {Richard}, F. and {Rocchi}, A. and {Rowan}, S. and {R{\"u}diger}, A. and {Santamar{\'\i}a}, L. and {Sassolas}, B. and {Sathyaprakash}, B. and {Schnabel}, R. and {Schwarz}, C. and {Seidel}, P. and {Sintes}, A. and {Somiya}, K. and {Speirits}, F. and {Strain}, K. and {Strigin}, S. and {Sutton}, P. and {Tarabrin}, S. and {Th{\"u}ring}, A. and {van den Brand}, J. and {van Veggel}, M. and {van den Broeck}, C. and {Vecchio}, A. and {Veitch}, J. and {Vetrano}, F. and {Vicere}, A. and {Vyatchanin}, S. and {Willke}, B. and {Woan}, G. and {Yamamoto}, K.},
        title = "{Sensitivity studies for third-generation gravitational wave observatories}",
      journal = {Classical and Quantum Gravity},
         year = 2011,
        month = may,
       volume = {28},
       number = {9},
          eid = {094013},
        pages = {094013},
          doi = {10.1088/0264-9381/28/9/094013},
archivePrefix = {arXiv},
       eprint = {1012.0908},
 primaryClass = {gr-qc},
       adsurl = {https://ui.adsabs.harvard.edu/abs/2011CQGra..28i4013H}
}

@ARTICLE{Bastian:2021,
       author = {{Bastian}, Niels-Uwe F.},
        title = "{Phenomenological quark-hadron equations of state with first-order phase transitions for astrophysical applications}",
      journal = {\prd},
         year = 2021,
        month = jan,
       volume = {103},
       number = {2},
          eid = {023001},
        pages = {023001},
          doi = {10.1103/PhysRevD.103.023001},
archivePrefix = {arXiv},
       eprint = {2009.10846},
 primaryClass = {nucl-th},
       adsurl = {https://ui.adsabs.harvard.edu/abs/2021PhRvD.103b3001B}
}

@ARTICLE{Nakazato13,
       author = {{Nakazato}, Ken'ichiro and {Sumiyoshi}, Kohsuke and {Yamada}, Shoichi},
        title = "{Stellar core collapse with hadron-quark phase transition}",
      journal = {\aap},
         year = 2013,
        month = oct,
       volume = {558},
          eid = {A50},
        pages = {A50},
          doi = {10.1051/0004-6361/201322231},
archivePrefix = {arXiv},
       eprint = {1309.3383},
 primaryClass = {astro-ph.HE},
       adsurl = {https://ui.adsabs.harvard.edu/abs/2013A&A...558A..50N}
}

@article{sotani_eos,
  title = {Universal relation for supernova gravitational waves},
  author = {Sotani, Hajime and Takiwaki, Tomoya and Togashi, Hajime},
  journal = {Phys. Rev. D},
  volume = {104},
  issue = {12},
  pages = {123009},
  numpages = {10},
  year = {2021},
  month = {Dec},
  publisher = {American Physical Society},
  doi = {10.1103/PhysRevD.104.123009},
  url = {https://link.aps.org/doi/10.1103/PhysRevD.104.123009}
}

@ARTICLE{sotani17,
   author = {{Sotani}, H. and {Kuroda}, T. and {Takiwaki}, T. and {Kotake}, K.
	},
    title = "{Probing mass-radius relation of protoneutron stars from gravitational-wave asteroseismology}",
  journal = {\prd},
archivePrefix = "arXiv",
   eprint = {1708.03738},
 primaryClass = "astro-ph.HE",
     year = 2017,
    month = sep,
   volume = 96,
   number = 6,
      eid = {063005},
    pages = {063005},
      doi = {10.1103/PhysRevD.96.063005},
   adsurl = {http://adsabs.harvard.edu/abs/2017PhRvD..96f3005S}
}

@ARTICLE{sotani19,
       author = {{Sotani}, Hajime and {Sumiyoshi}, Kohsuke},
        title = "{Determination of properties of protoneutron stars toward black hole formation via gravitational wave observations}",
      journal = {\prd},
         year = 2019,
        month = oct,
       volume = {100},
       number = {8},
          eid = {083008},
        pages = {083008},
          doi = {10.1103/PhysRevD.100.083008},
archivePrefix = {arXiv},
       eprint = {1909.11816},
 primaryClass = {astro-ph.HE},
       adsurl = {https://ui.adsabs.harvard.edu/abs/2019PhRvD.100h3008S}
}

@ARTICLE{abbott18det,
   author = {{Abbott}, B.~P. and {Abbott}, R. and {Abbott}, T.~D. and {Abernathy}, M.~R. and 
	{Acernese}, F. and {Ackley}, K. and {Adams}, C. and {Adams}, T. and 
	{Addesso}, P. and {Adhikari}, R.~X. and et al.},
    title = "{Prospects for observing and localizing gravitational-wave transients with Advanced LIGO, Advanced Virgo and KAGRA}",
  journal = {Living Reviews in Relativity},
archivePrefix = "arXiv",
   eprint = {1304.0670},
 primaryClass = "gr-qc",
     year = 2018,
    month = apr,
   volume = 21,
      eid = {3},
    pages = {3},
      doi = {10.1007/s41114-018-0012-9},
   adsurl = {https://ui.adsabs.harvard.edu/abs/2018LRR....21....3A}
}

@ARTICLE{Fischer18,
       author = {{Fischer}, Tobias and {Bastian}, Niels-Uwe F. and {Wu}, Meng-Ru and {Baklanov}, Petr and {Sorokina}, Elena and {Blinnikov}, Sergei and {Typel}, Stefan and {Kl{\"a}hn}, Thomas and {Blaschke}, David B.},
        title = "{Quark deconfinement as a supernova explosion engine for massive blue supergiant stars}",
      journal = {Nature Astronomy},
         year = 2018,
        month = oct,
       volume = {2},
        pages = {980-986},
          doi = {10.1038/s41550-018-0583-0},
archivePrefix = {arXiv},
       eprint = {1712.08788},
 primaryClass = {astro-ph.HE},
       adsurl = {https://ui.adsabs.harvard.edu/abs/2018NatAs...2..980F}
}

@ARTICLE{Shibata_JFBD_14,
       author = {{Shibata}, Masaru and {Taniguchi}, Keisuke and {Okawa}, Hirotada and {Buonanno}, Alessandra},
        title = "{Coalescence of binary neutron stars in a scalar-tensor theory of gravity}",
      journal = {\prd},
         year = 2014,
        month = apr,
       volume = {89},
       number = {8},
          eid = {084005},
        pages = {084005},
          doi = {10.1103/PhysRevD.89.084005},
archivePrefix = {arXiv},
       eprint = {1310.0627},
 primaryClass = {gr-qc},
       adsurl = {https://ui.adsabs.harvard.edu/abs/2014PhRvD..89h4005S}
}

@ARTICLE{Kotake18,
       author = {{Kotake}, Kei and {Takiwaki}, Tomoya and {Fischer}, Tobias and
         {Nakamura}, Ko and {Mart{\'\i}nez-Pinedo}, Gabriel},
        title = "{Impact of Neutrino Opacities on Core-collapse Supernova Simulations}",
      journal = {\apj},
         year = "2018",
        month = "Feb",
       volume = {853},
       number = {2},
          eid = {170},
        pages = {170},
          doi = {10.3847/1538-4357/aaa716},
archivePrefix = {arXiv},
       eprint = {1801.02703},
 primaryClass = {astro-ph.HE},
       adsurl = {https://ui.adsabs.harvard.edu/abs/2018ApJ...853..170K}
}

@ARTICLE{Nakamura22,
       author = {{Nakamura}, Ko and {Takiwaki}, Tomoya and {Kotake}, Kei},
        title = "{Three-dimensional simulation of a core-collapse supernova for a binary star progenitor of SN 1987A}",
      journal = {\mnras},
         year = 2022,
        month = aug,
       volume = {514},
       number = {3},
        pages = {3941-3952},
          doi = {10.1093/mnras/stac1586},
archivePrefix = {arXiv},
       eprint = {2202.06295},
 primaryClass = {astro-ph.HE},
       adsurl = {https://ui.adsabs.harvard.edu/abs/2022MNRAS.514.3941N}
}

@ARTICLE{Ott09,
   author = {{Ott}, C.~D},
    title = "{TOPICAL REVIEW:  The gravitational-wave signature of core-collapse supernovae}",
  journal = {Classical and Quantum Gravity},
archivePrefix = "arXiv",
   eprint = {0809.0695},
     year = 2009,
    month = mar,
   volume = 26,
   number = 6,
    pages = {063001},
      doi = {10.1088/0264-9381/26/6/063001},
   adsurl = {http://ads.nao.ac.jp/abs/2009CQGra..26f3001O}
}

@ARTICLE{Kokkotas99,
       author = {{Kokkotas}, Kostas D. and {Schmidt}, Bernd G.},
        title = "{Quasi-Normal Modes of Stars and Black Holes}",
      journal = {Living Reviews in Relativity},
         year = 1999,
        month = dec,
       volume = {2},
       number = {1},
          eid = {2},
        pages = {2},
          doi = {10.12942/lrr-1999-2},
archivePrefix = {arXiv},
       eprint = {gr-qc/9909058},
 primaryClass = {gr-qc},
       adsurl = {https://ui.adsabs.harvard.edu/abs/1999LRR.....2....2K}
}

@ARTICLE{Lella26,
       author = {{Lella}, Alessandro and {Lucente}, Giuseppe and {Kresse}, Daniel and {Glas}, Robert and {Janka}, H.-Thomas and {Mirizzi}, Alessandro},
        title = "{Gravitational-Wave Signals for Supernova Explosions of Three-Dimensional Progenitors}",
      journal = {arXiv e-prints},
         year = 2026,
        month = feb,
          eid = {arXiv:2602.02651},
        pages = {arXiv:2602.02651},
          doi = {10.48550/arXiv.2602.02651},
archivePrefix = {arXiv},
       eprint = {2602.02651},
 primaryClass = {astro-ph.HE},
       adsurl = {https://ui.adsabs.harvard.edu/abs/2026arXiv260202651L}
}

@ARTICLE{Vartanyan23,
       author = {{Vartanyan}, David and {Burrows}, Adam and {Wang}, Tianshu and {Coleman}, Matthew S.~B. and {White}, Christopher J.},
        title = "{Gravitational-wave signature of core-collapse supernovae}",
      journal = {\prd},
         year = 2023,
        month = may,
       volume = {107},
       number = {10},
          eid = {103015},
        pages = {103015},
          doi = {10.1103/PhysRevD.107.103015},
archivePrefix = {arXiv},
       eprint = {2302.07092},
 primaryClass = {astro-ph.HE},
       adsurl = {https://ui.adsabs.harvard.edu/abs/2023PhRvD.107j3015V}
}

@ARTICLE{Powell21,
       author = {{Powell}, Jade and {M{\"u}ller}, Bernhard and {Heger}, Alexander},
        title = "{The final core collapse of pulsational pair instability supernovae}",
      journal = {\mnras},
         year = 2021,
        month = may,
       volume = {503},
       number = {2},
        pages = {2108-2122},
          doi = {10.1093/mnras/stab614},
archivePrefix = {arXiv},
       eprint = {2101.06889},
 primaryClass = {astro-ph.HE},
       adsurl = {https://ui.adsabs.harvard.edu/abs/2021MNRAS.503.2108P}
}

@ARTICLE{Fischer14,
   author = {{Fischer}, T. and {Hempel}, M. and {Sagert}, I. and {Suwa}, Y. and 
	{Schaffner-Bielich}, J.},
    title = "{Symmetry energy impact in simulations of core-collapse supernovae}",
  journal = {European Physical Journal A},
archivePrefix = "arXiv",
   eprint = {1307.6190},
 primaryClass = "astro-ph.HE",
     year = 2014,
    month = feb,
   volume = 50,
      eid = {46},
    pages = {46},
      doi = {10.1140/epja/i2014-14046-5},
   adsurl = {http://cdsads.u-strasbg.fr/abs/2014EPJA...50...46F}
}

@ARTICLE{fryer11,
   author = {{Fryer}, C.~L. and {New}, K.~C.~B.},
    title = "{Gravitational Waves from Gravitational Collapse}",
  journal = {Living Reviews in Relativity},
     year = 2011,
    month = jan,
   volume = 14,
    pages = {1-+},
   adsurl = {http://ads.nao.ac.jp/abs/2011LRR....14....1F}
}

@ARTICLE{Zwerger97,
       author = {{Zwerger}, T. and {M\"uller}, E.},
        title = "{Dynamics and gravitational wave signature of axisymmetric rotational core collapse.}",
      journal = {\aap},
         year = 1997,
        month = apr,
       volume = {320},
        pages = {209-227},
       adsurl = {https://ui.adsabs.harvard.edu/abs/1997A&A...320..209Z}
}

@ARTICLE{shibagaki26,
       author = {{Shibagaki}, Shota and {Takiwaki}, Tomoya and {Kotake}, Kei and {Kuroda}, Takami and {Fischer}, Tobias},
        title = "{Circular polarization of gravitational waves from magnetorotational supernovae}",
      journal = {\aap},
         year = 2026,
        month = jul,
       volume = {711},
          eid = {L6},
        pages = {L6},
          doi = {10.1051/0004-6361/202659979},
archivePrefix = {arXiv},
       eprint = {2603.20473},
 primaryClass = {astro-ph.HE},
       adsurl = {https://ui.adsabs.harvard.edu/abs/2026A&A...711L...6S}
}

@article{Abdikamalov14,
  title = {Measuring the angular momentum distribution in core-collapse supernova progenitors with gravitational waves},
  author = {Abdikamalov, Ernazar and Gossan, Sarah and DeMaio, Alexandra M. and Ott, Christian D.},
  journal = {Phys. Rev. D},
  volume = {90},
  issue = {4},
  pages = {044001},
  numpages = {25},
  year = {2014},
  month = {Aug},
  publisher = {American Physical Society},
  doi = {10.1103/PhysRevD.90.044001},
  url = {https://link.aps.org/doi/10.1103/PhysRevD.90.044001}
}

@ARTICLE{Andresen17,
   author = {{Andresen}, H. and {M{\"u}ller}, B. and {M{\"u}ller}, E. and 
	{Janka}, H.-T.},
    title = "{Gravitational wave signals from 3D neutrino hydrodynamics simulations of core-collapse supernovae}",
  journal = {\mnras},
archivePrefix = "arXiv",
   eprint = {1607.05199},
 primaryClass = "astro-ph.HE",
     year = 2017,
    month = jun,
   volume = 468,
    pages = {2032-2051},
      doi = {10.1093/mnras/stx618},
   adsurl = {http://cdsads.u-strasbg.fr/abs/2017MNRAS.468.2032A}
}

@ARTICLE{eggenberger21,
       author = {{Eggenberger Andersen}, Oliver and {Zha}, Shuai and {da Silva Schneider}, Andr{\'e} and {Betranhandy}, Aurore and {Couch}, Sean M. and {O'Connor}, Evan P.},
        title = "{Equation-of-state Dependence of Gravitational Waves in Core-collapse Supernovae}",
      journal = {\apj},
         year = 2021,
        month = dec,
       volume = {923},
       number = {2},
          eid = {201},
        pages = {201},
          doi = {10.3847/1538-4357/ac294c},
archivePrefix = {arXiv},
       eprint = {2106.09734},
 primaryClass = {astro-ph.HE},
       adsurl = {https://ui.adsabs.harvard.edu/abs/2021ApJ...923..201E}
}

@ARTICLE{Andresen21,
       author = {{Andresen}, H. and {Glas}, R. and {Janka}, H. -Th},
        title = "{Gravitational-wave signals from 3D supernova simulations with different neutrino-transport methods}",
      journal = {\mnras},
         year = 2021,
        month = may,
       volume = {503},
       number = {3},
        pages = {3552-3567},
          doi = {10.1093/mnras/stab675},
archivePrefix = {arXiv},
       eprint = {2011.10499},
 primaryClass = {astro-ph.HE},
       adsurl = {https://ui.adsabs.harvard.edu/abs/2021MNRAS.503.3552A}
}

@ARTICLE{Blondin07_nat,
   author = {{Blondin}, J.~M. and {Mezzacappa}, A.},
    title = "{Pulsar spins from an instability in the accretion shock of supernovae}",
  journal = {\nat},
   eprint = {arXiv:astro-ph/0611680},
     year = 2007,
    month = jan,
   volume = 445,
    pages = {58-60},
      doi = {10.1038/nature05428},
   adsurl = {http://ads.nao.ac.jp/abs/2007Natur.445...58B}
}

@ARTICLE{Buras06a,
       author = {{Buras}, R. and {Rampp}, M. and {Janka}, H.-Th. and {Kifonidis}, K.},
        title = "{Two-dimensional hydrodynamic core-collapse supernova simulations with spectral neutrino transport. I. Numerical method and results for a 15 $M_{odot}$ star}",
      journal = {\aap},
         year = 2006,
        month = mar,
       volume = {447},
       number = {3},
        pages = {1049-1092},
          doi = {10.1051/0004-6361:20053783},
archivePrefix = {arXiv},
       eprint = {astro-ph/0507135},
 primaryClass = {astro-ph},
       adsurl = {https://ui.adsabs.harvard.edu/abs/2006A&A...447.1049B}
}

@ARTICLE{Mezzacappa23,
       author = {{Mezzacappa}, Anthony and {Marronetti}, Pedro and {Landfield}, Ryan E. and {Lentz}, Eric J. and {Murphy}, R. Daniel and {Raphael Hix}, W. and {Harris}, J. Austin and {Bruenn}, Stephen W. and {Blondin}, John M. and {Bronson Messer}, O.~E. and {Casanova}, Jordi and {Kronzer}, Luke L.},
        title = "{Core collapse supernova gravitational wave emission for progenitors of 9.6, 15, and 25$M_{\odot}$}",
      journal = {\prd},
         year = 2023,
        month = feb,
       volume = {107},
       number = {4},
          eid = {043008},
        pages = {043008},
          doi = {10.1103/PhysRevD.107.043008},
archivePrefix = {arXiv},
       eprint = {2208.10643},
 primaryClass = {astro-ph.SR},
       adsurl = {https://ui.adsabs.harvard.edu/abs/2023PhRvD.107d3008M}
}

@ARTICLE{Buras06b,
   author = {{Buras}, R. and {Janka}, H.-T. and {Rampp}, M. and {Kifonidis}, K.
	},
    title = "{Two-dimensional hydrodynamic core-collapse supernova simulations with spectral neutrino transport. II. Models for different progenitor stars}",
  journal = {\aap},
   eprint = {astro-ph/0512189},
     year = 2006,
    month = oct,
   volume = 457,
    pages = {281-308},
      doi = {10.1051/0004-6361:20054654},
   adsurl = {http://ads.nao.ac.jp/abs/2006A%26A...457..281B}
}

@ARTICLE{bugli23,
       author = {{Bugli}, M. and {Guilet}, J. and {Foglizzo}, T. and {Obergaulinger}, M.},
        title = "{Three-dimensional core-collapse supernovae with complex magnetic structures - II. Rotational instabilities and multimessenger signatures}",
      journal = {\mnras},
         year = 2023,
        month = apr,
       volume = {520},
       number = {4},
        pages = {5622-5634},
          doi = {10.1093/mnras/stad496},
archivePrefix = {arXiv},
       eprint = {2210.05012},
 primaryClass = {astro-ph.HE},
       adsurl = {https://ui.adsabs.harvard.edu/abs/2023MNRAS.520.5622B}
}

@ARTICLE{Chan&Muller18,
   author = {{Chan}, C. and {M{\"u}ller}, B. and {Heger}, A. and {Pakmor}, R. and 
	{Springel}, V.},
    title = "{Black Hole Formation and Fallback during the Supernova Explosion of a 40 M $_{\odot}$ Star}",
  journal = {\apjl},
archivePrefix = "arXiv",
   eprint = {1710.00838},
 primaryClass = "astro-ph.SR",
     year = 2018,
    month = jan,
   volume = 852,
      eid = {L19},
    pages = {L19},
      doi = {10.3847/2041-8213/aaa28c},
   adsurl = {http://cdsads.u-strasbg.fr/abs/2018ApJ...852L..19C}
}

@ARTICLE{Dimmelmeier02B,
   author = {{Dimmelmeier}, H. and {Font}, J.~A. and {M{\"u}ller}, E.},
    title = "{Relativistic simulations of rotational core collapse II. Collapse dynamics and gravitational radiation}",
  journal = {\aap},
   eprint = {astro-ph/0204289},
     year = 2002,
    month = oct,
   volume = 393,
    pages = {523-542},
      doi = {10.1051/0004-6361:20021053},
   adsurl = {http://ads.nao.ac.jp/abs/2002A%26A...393..523D}
}

@ARTICLE{Dimmelmeier08,
   author = {{Dimmelmeier}, H. and {Ott}, C.~D. and {Marek}, A. and {Janka}, H.-T.
	},
    title = "{Gravitational wave burst signal from core collapse of rotating stars}",
  journal = {\prd},
archivePrefix = "arXiv",
   eprint = {0806.4953},
     year = 2008,
    month = sep,
   volume = 78,
   number = 6,
      eid = {064056},
    pages = {064056},
      doi = {10.1103/PhysRevD.78.064056},
   adsurl = {http://ads.nao.ac.jp/abs/2008PhRvD..78f4056D}
}

@ARTICLE{Foglizzo06,
   author = {{Foglizzo}, T. and {Scheck}, L. and {Janka}, H.-T.},
    title = "{Neutrino-driven Convection versus Advection in Core-Collapse Supernovae}",
  journal = {\apj},
   eprint = {arXiv:astro-ph/0507636},
     year = 2006,
    month = dec,
   volume = 652,
    pages = {1436-1450},
      doi = {10.1086/508443},
   adsurl = {http://ads.nao.ac.jp/abs/2006ApJ...652.1436F}
}

@ARTICLE{Foglizzo07,
   author = {{Foglizzo}, T. and {Galletti}, P. and {Scheck}, L. and {Janka}, H.-T.
	},
    title = "{Instability of a Stalled Accretion Shock: Evidence for the Advective-Acoustic Cycle}",
  journal = {\apj},
   eprint = {arXiv:astro-ph/0606640},
     year = 2007,
    month = jan,
   volume = 654,
    pages = {1006-1021},
      doi = {10.1086/509612},
   adsurl = {http://ads.nao.ac.jp/abs/2007ApJ...654.1006F}
}

@ARTICLE{Foglizzo09,
   author = {{Foglizzo}, T.},
    title = "{A Simple Toy Model of the Advective-Acoustic Instability. I. Perturbative Approach}",
  journal = {\apj},
archivePrefix = "arXiv",
   eprint = {0809.2302},
     year = 2009,
    month = apr,
   volume = 694,
    pages = {820-832},
      doi = {10.1088/0004-637X/694/2/820},
   adsurl = {http://ads.nao.ac.jp/abs/2009ApJ...694..820F}
}

@ARTICLE{Foglizzo12,
   author = {{Foglizzo}, T. and {Masset}, F. and {Guilet}, J. and {Durand}, G.
	},
    title = "{Shallow Water Analogue of the Standing Accretion Shock Instability: Experimental Demonstration and a Two-Dimensional Model}",
  journal = {Physical Review Letters},
archivePrefix = "arXiv",
   eprint = {1112.3448},
 primaryClass = "astro-ph.HE",
     year = 2012,
    month = feb,
   volume = 108,
   number = 5,
      eid = {051103},
    pages = {051103},
      doi = {10.1103/PhysRevLett.108.051103},
   adsurl = {http://ads.nao.ac.jp/abs/2012PhRvL.108e1103F}
}

@ARTICLE{Foglizzo15,
       author = {{Foglizzo}, Thierry and {Kazeroni}, R{\'e}mi and
         {Guilet}, J{\'e}r{\^o}me and {Masset}, Fr{\'e}d{\'e}ric and
         {Gonz{\'a}lez}, Matthias and {Krueger}, Brendan K. and
         {Novak}, J{\'e}r{\^o}me and {Oertel}, Micaela and
         {Margueron}, J{\'e}r{\^o}me and {Faure}, Julien and {Martin}, No{\"e}l and
         {Blottiau}, Patrick and {Peres}, Bruno and {Durand}, Gilles},
        title = "{The Explosion Mechanism of Core-Collapse Supernovae: Progress in Supernova Theory and Experiments}",
      journal = {\pasa},
         year = "2015",
        month = "Mar",
       volume = {32},
          eid = {e009},
        pages = {e009},
          doi = {10.1017/pasa.2015.9},
archivePrefix = {arXiv},
       eprint = {1501.01334},
 primaryClass = {astro-ph.HE},
       adsurl = {https://ui.adsabs.harvard.edu/abs/2015PASA...32....9F}
}

@ARTICLE{nemo,
       author = {{Ackley}, K. and {Adya}, V.~B. and {Agrawal}, P. and {Altin}, P. and {Ashton}, G. and {Bailes}, M. and {Baltinas}, E. and {Barbuio}, A. and {Beniwal}, D. and {Blair}, C. and {Blair}, D. and {Bolingbroke}, G.~N. and {Bossilkov}, V. and {Shachar Boublil}, S. and {Brown}, D.~D. and {Burridge}, B.~J. and {Calderon Bustillo}, J. and {Cameron}, J. and {Tuong Cao}, H. and {Carlin}, J.~B. and {Chang}, S. and {Charlton}, P. and {Chatterjee}, C. and {Chattopadhyay}, D. and {Chen}, X. and {Chi}, J. and {Chow}, J. and {Chu}, Q. and {Ciobanu}, A. and {Clarke}, T. and {Clearwater}, P. and {Cooke}, J. and {Coward}, D. and {Crisp}, H. and {Dattatri}, R.~J. and {Deller}, A.~T. and {Dobie}, D.~A. and {Dunn}, L. and {Easter}, P.~J. and {Eichholz}, J. and {Evans}, R. and {Flynn}, C. and {Foran}, G. and {Forsyth}, P. and {Gai}, Y. and {Galaudage}, S. and {Galloway}, D.~K. and {Gendre}, B. and {Goncharov}, B. and {Goode}, S. and {Gozzard}, D. and {Grace}, B. and {Graham}, A.~W. and {Heger}, A. and {Hernandez Vivanco}, F. and {Hirai}, R. and {Holland}, N.~A. and {Holmes}, Z.~J. and {Howard}, E. and {Howell}, E. and {Howitt}, G. and {H{\"u}bner}, M.~T. and {Hurley}, J. and {Ingram}, C. and {Jaberian Hamedan}, V. and {Jenner}, K. and {Ju}, L. and {Kapasi}, D.~P. and {Kaur}, T. and {Kijbunchoo}, N. and {Kovalam}, M. and {Kumar Choudhary}, R. and {Lasky}, P.~D. and {Lau}, M.~Y.~M. and {Leung}, J. and {Liu}, J. and {Loh}, K. and {Mailvagan}, A. and {Mandel}, I. and {McCann}, J.~J. and {McClelland}, D.~E. and {McKenzie}, K. and {McManus}, D. and {McRae}, T. and {Melatos}, A. and {Meyers}, P. and {Middleton}, H. and {Miles}, M.~T. and {Millhouse}, M. and {Lun Mong}, Y. and {Mueller}, B. and {Munch}, J. and {Musiov}, J. and {Muusse}, S. and {Nathan}, R.~S. and {Naveh}, Y. and {Neijssel}, C. and {Neil}, B. and {Ng}, S.~W.~S. and {Oloworaran}, V. and {Ottaway}, D.~J. and {Page}, M. and {Pan}, J. and {Pathak}, M. and {Payne}, E. and {Powell}, J. and {Pritchard}, J. and {Puckridge}, E. and {Raidani}, A. and {Rallabhandi}, V. and {Reardon}, D. and {Riley}, J.~A. and {Roberts}, L. and {Romero-Shaw}, I.~M. and {Roocke}, T.~J. and {Rowell}, G. and {Sahu}, N. and {Sarin}, N. and {Sarre}, L. and {Sattari}, H. and {Schiworski}, M. and {Scott}, S.~M. and {Sengar}, R. and {Shaddock}, D. and {Shannon}, R. and {SHI}, J. and {Sibley}, P. and {Slagmolen}, B.~J.~J. and {Slaven-Blair}, T. and {Smith}, R.~J.~E. and {Spollard}, J. and {Steed}, L. and {Strang}, L. and {Sun}, H. and {Sunderland}, A. and {Suvorova}, S. and {Talbot}, C. and {Thrane}, E. and {T{\"o}yr{\"a}}, D. and {Trahanas}, P. and {Vajpeyi}, A. and {van Heijningen}, J.~V. and {Vargas}, A.~F. and {Veitch}, P.~J. and {Vigna-Gomez}, A. and {Wade}, A. and {Walker}, K. and {Wang}, Z. and {Ward}, R.~L. and {Ward}, K. and {Webb}, S. and {Wen}, L. and {Wette}, K. and {Wilcox}, R. and {Winterflood}, J. and {Wolf}, C. and {Wu}, B. and {Jet Yap}, M. and {You}, Z. and {Yu}, H. and {Zhang}, J. and {Zhang}, J. and {Zhao}, C. and {Zhu}, X.},
        title = "{Neutron Star Extreme Matter Observatory: A kilohertz-band gravitational-wave detector in the global network}",
      journal = {\pasa},
         year = 2020,
        month = nov,
       volume = {37},
          eid = {e047},
        pages = {e047},
          doi = {10.1017/pasa.2020.39},
archivePrefix = {arXiv},
       eprint = {2007.03128},
 primaryClass = {astro-ph.HE},
       adsurl = {https://ui.adsabs.harvard.edu/abs/2020PASA...37...47A}
}

@ARTICLE{jade24,
       author = {{Powell}, Jade and {Iess}, Alberto and {Llorens-Monteagudo}, Miquel and {Obergaulinger}, Martin and {M{\"u}ller}, Bernhard and {Torres-Forn{\'e}}, Alejandro and {Cuoco}, Elena and {Font}, Jos{\'e} A.},
        title = "{Determining the core-collapse supernova explosion mechanism with current and future gravitational-wave observatories}",
      journal = {\prd},
         year = 2024,
        month = mar,
       volume = {109},
       number = {6},
          eid = {063019},
        pages = {063019},
          doi = {10.1103/PhysRevD.109.063019},
archivePrefix = {arXiv},
       eprint = {2311.18221},
 primaryClass = {astro-ph.HE},
       adsurl = {https://ui.adsabs.harvard.edu/abs/2024PhRvD.109f3019P}
}

@ARTICLE{ernazar2020,
       author = {{Abdikamalov}, Ernazar and {Pagliaroli}, Giulia and {Radice}, David},
        title = "{Gravitational Waves from Core-Collapse Supernovae}",
      journal = {arXiv e-prints},
         year = 2020,
        month = oct,
          eid = {arXiv:2010.04356},
        pages = {arXiv:2010.04356},
          doi = {10.48550/arXiv.2010.04356},
archivePrefix = {arXiv},
       eprint = {2010.04356},
 primaryClass = {astro-ph.SR},
       adsurl = {https://ui.adsabs.harvard.edu/abs/2020arXiv201004356A}
}

@ARTICLE{bernhard2026,
       author = {{M{\"u}ller}, Bernhard},
        title = "{Core-Collapse Supernovae and their Gravitational Wave Signals: The Status of Theory and Modeling}",
      journal = {arXiv e-prints},
         year = 2026,
        month = mar,
          eid = {arXiv:2603.24243},
        pages = {arXiv:2603.24243},
          doi = {10.48550/arXiv.2603.24243},
archivePrefix = {arXiv},
       eprint = {2603.24243},
 primaryClass = {astro-ph.HE},
       adsurl = {https://ui.adsabs.harvard.edu/abs/2026arXiv260324243M}
}

@ARTICLE{Hayama18,
       author = {{Hayama}, Kazuhiro and {Kuroda}, Takami and {Kotake}, Kei and {Takiwaki}, Tomoya},
        title = "{Circular polarization of gravitational waves from non-rotating supernova cores: a new probe into the pre-explosion hydrodynamics}",
      journal = {\mnras},
         year = 2018,
        month = jun,
       volume = {477},
       number = {1},
        pages = {L96-L100},
          doi = {10.1093/mnrasl/sly055},
archivePrefix = {arXiv},
       eprint = {1802.03842},
 primaryClass = {astro-ph.HE},
       adsurl = {https://ui.adsabs.harvard.edu/abs/2018MNRAS.477L..96H}
}

@ARTICLE{Kawahara18,
       author = {{Kawahara}, Hajime and {Kuroda}, Takami and {Takiwaki}, Tomoya and {Hayama}, Kazuhiro and {Kotake}, Kei},
        title = "{A Linear and Quadratic Time-Frequency Analysis of Gravitational Waves from Core-collapse Supernovae}",
      journal = {\apj},
         year = 2018,
        month = nov,
       volume = {867},
       number = {2},
          eid = {126},
        pages = {126},
          doi = {10.3847/1538-4357/aae57b},
archivePrefix = {arXiv},
       eprint = {1810.00334},
 primaryClass = {astro-ph.HE},
       adsurl = {https://ui.adsabs.harvard.edu/abs/2018ApJ...867..126K}
}

@ARTICLE{Kotake06,
   author = {{Kotake}, K. and {Sato}, K. and {Takahashi}, K.},
    title = "{Explosion mechanism, neutrino burst and gravitational wave in core-collapse supernovae}",
  journal = {Reports on Progress in Physics},
   eprint = {arXiv:astro-ph/0509456},
     year = 2006,
    month = apr,
   volume = 69,
    pages = {971-1143},
      doi = {10.1088/0034-4885/69/4/R03},
   adsurl = {http://ads.nao.ac.jp/abs/2006RPPh...69..971K}
}

@ARTICLE{Kotake12_ptep,
   author = {{Kotake}, K. and {Sumiyoshi}, K. and {Yamada}, S. and {Takiwaki}, T. and 
	{Kuroda}, T. and {Suwa}, Y. and {Nagakura}, H.},
    title = "{Core-collapse supernovae as supercomputing science: A status report toward six-dimensional simulations with exact Boltzmann neutrino transport in full general relativity}",
  journal = {Progress of Theoretical and Experimental Physics},
archivePrefix = "arXiv",
   eprint = {1205.6284},
 primaryClass = "astro-ph.HE",
     year = 2012,
    month = aug,
   volume = 2012,
   number = 1,
      eid = {01A301},
    pages = {010000},
      doi = {10.1093/ptep/pts009},
   adsurl = {http://ads.nao.ac.jp/abs/2012PTEP.2012aA301K}
}

@ARTICLE{Kotake13,
   author = {{Kotake}, K.},
    title = "{Multiple physical elements to determine the gravitational-wave signatures of core-collapse supernovae}",
  journal = {Comptes Rendus Physique},
     year = 2013,
archivePrefix = "arXiv",
   eprint = {1110.5107},
    month = apr,
   volume = 14,
    pages = {318-351},
      doi = {10.1016/j.crhy.2013.01.008},
   adsurl = {http://adsabs.harvard.edu/abs/2013CRPhy..14..318K}
}

@ARTICLE{GR_Test_berti15,
       author = {{Berti}, Emanuele and {Barausse}, Enrico and {Cardoso}, Vitor and {Gualtieri}, Leonardo and {Pani}, Paolo and {Sperhake}, Ulrich and {Stein}, Leo C. and {Wex}, Norbert and {Yagi}, Kent and {Baker}, Tessa and {Burgess}, C.~P. and {Coelho}, Fl{\'a}vio S. and {Doneva}, Daniela and {Felice}, Antonio De and {Ferreira}, Pedro G. and {Freire}, Paulo C.~C. and {Healy}, James and {Herdeiro}, Carlos and {Horbatsch}, Michael and {Kleihaus}, Burkhard and {Klein}, Antoine and {Kokkotas}, Kostas and {Kunz}, Jutta and {Laguna}, Pablo and {Lang}, Ryan N. and {Li}, Tjonnie G.~F. and {Littenberg}, Tyson and {Matas}, Andrew and {Mirshekari}, Saeed and {Okawa}, Hirotada and {Radu}, Eugen and {O'Shaughnessy}, Richard and {Sathyaprakash}, Bangalore S. and {Broeck}, Chris Van Den and {Winther}, Hans A. and {Witek}, Helvi and {Aghili}, Mir Emad and {Alsing}, Justin and {Bolen}, Brett and {Bombelli}, Luca and {Caudill}, Sarah and {Chen}, Liang and {Degollado}, Juan Carlos and {Fujita}, Ryuichi and {Gao}, Caixia and {Gerosa}, Davide and {Kamali}, Saeed and {Silva}, Hector O. and {Rosa}, Jo{\~a}o G. and {Sadeghian}, Laleh and {Sampaio}, Marco and {Sotani}, Hajime and {Zilhao}, Miguel},
        title = "{Testing general relativity with present and future astrophysical observations}",
      journal = {Classical and Quantum Gravity},
         year = 2015,
        month = dec,
       volume = {32},
       number = {24},
          eid = {243001},
        pages = {243001},
          doi = {10.1088/0264-9381/32/24/24300110.48550/arXiv.1501.07274},
archivePrefix = {arXiv},
       eprint = {1501.07274},
 primaryClass = {gr-qc},
       adsurl = {https://ui.adsabs.harvard.edu/abs/2015CQGra..32x3001B}
}

@ARTICLE{GR_Test_Psaltis08,
       author = {{Psaltis}, Dimitrios},
        title = "{Probes and Tests of Strong-Field Gravity with Observations in the Electromagnetic Spectrum}",
      journal = {Living Reviews in Relativity},
         year = 2008,
        month = nov,
       volume = {11},
       number = {1},
          eid = {9},
        pages = {9},
          doi = {10.12942/lrr-2008-910.48550/arXiv.0806.1531},
archivePrefix = {arXiv},
       eprint = {0806.1531},
 primaryClass = {astro-ph},
       adsurl = {https://ui.adsabs.harvard.edu/abs/2008LRR....11....9P}
}

@ARTICLE{GR_Test_Will14,
        author = {{Will}, Clifford M.},
        title = "{The Confrontation between General Relativity and Experiment}",
      journal = {Living Reviews in Relativity},
         year = 2014,
        month = dec,
       volume = {17},
       number = {1},
          eid = {4},
        pages = {4},
          doi = {10.12942/lrr-2014-410.48550/arXiv.1403.7377},
archivePrefix = {arXiv},
       eprint = {1403.7377},
 primaryClass = {gr-qc},
       adsurl = {https://ui.adsabs.harvard.edu/abs/2014LRR....17....4W}
}

@ARTICLE{Perlmutter99,
       author = {{Perlmutter}, S. and {Aldering}, G. and {Goldhaber}, G. and {Knop}, R.~A. and {Nugent}, P. and {Castro}, P.~G. and {Deustua}, S. and {Fabbro}, S. and {Goobar}, A. and {Groom}, D.~E. and {Hook}, I.~M. and {Kim}, A.~G. and {Kim}, M.~Y. and {Lee}, J.~C. and {Nunes}, N.~J. and {Pain}, R. and {Pennypacker}, C.~R. and {Quimby}, R. and {Lidman}, C. and {Ellis}, R.~S. and {Irwin}, M. and {McMahon}, R.~G. and {Ruiz-Lapuente}, P. and {Walton}, N. and {Schaefer}, B. and {Boyle}, B.~J. and {Filippenko}, A.~V. and {Matheson}, T. and {Fruchter}, A.~S. and {Panagia}, N. and {Newberg}, H.~J.~M. and {Couch}, W.~J. and {Project}, The Supernova Cosmology},
        title = "{Measurements of {\ensuremath{\Omega}} and {\ensuremath{\Lambda}} from 42 High-Redshift Supernovae}",
      journal = {\apj},
         year = 1999,
        month = jun,
       volume = {517},
       number = {2},
        pages = {565-586},
          doi = {10.1086/307221},
archivePrefix = {arXiv},
       eprint = {astro-ph/9812133},
 primaryClass = {astro-ph},
       adsurl = {https://ui.adsabs.harvard.edu/abs/1999ApJ...517..565P}
}

@ARTICLE{Komatsu7yrs,
       author = {{Komatsu}, E. and {Smith}, K.~M. and {Dunkley}, J. and {Bennett}, C.~L. and {Gold}, B. and {Hinshaw}, G. and {Jarosik}, N. and {Larson}, D. and {Nolta}, M.~R. and {Page}, L. and {Spergel}, D.~N. and {Halpern}, M. and {Hill}, R.~S. and {Kogut}, A. and {Limon}, M. and {Meyer}, S.~S. and {Odegard}, N. and {Tucker}, G.~S. and {Weiland}, J.~L. and {Wollack}, E. and {Wright}, E.~L.},
        title = "{Seven-year Wilkinson Microwave Anisotropy Probe (WMAP) Observations: Cosmological Interpretation}",
      journal = {\apjs},
         year = 2011,
        month = feb,
       volume = {192},
       number = {2},
          eid = {18},
        pages = {18},
          doi = {10.1088/0067-0049/192/2/18},
archivePrefix = {arXiv},
       eprint = {1001.4538},
 primaryClass = {astro-ph.CO},
       adsurl = {https://ui.adsabs.harvard.edu/abs/2011ApJS..192...18K}
}

@ARTICLE{Abbott19,
       author = {{Abbott}, T.~M.~C. and {Allam}, S. and {Andersen}, P. and {Angus}, C. and {Asorey}, J. and {Avelino}, A. and {Avila}, S. and {Bassett}, B.~A. and {Bechtol}, K. and {Bernstein}, G.~M. and {Bertin}, E. and {Brooks}, D. and {Brout}, D. and {Brown}, P. and {Burke}, D.~L. and {Calcino}, J. and {Carnero Rosell}, A. and {Carollo}, D. and {Carrasco Kind}, M. and {Carretero}, J. and {Casas}, R. and {Castander}, F.~J. and {Cawthon}, R. and {Challis}, P. and {Childress}, M. and {Clocchiatti}, A. and {Cunha}, C.~E. and {D'Andrea}, C.~B. and {da Costa}, L.~N. and {Davis}, C. and {Davis}, T.~M. and {De Vicente}, J. and {DePoy}, D.~L. and {Desai}, S. and {Diehl}, H.~T. and {Doel}, P. and {Drlica-Wagner}, A. and {Eifler}, T.~F. and {Evrard}, A.~E. and {Fernandez}, E. and {Filippenko}, A.~V. and {Finley}, D.~A. and {Flaugher}, B. and {Foley}, R.~J. and {Fosalba}, P. and {Frieman}, J. and {Galbany}, L. and {Garc{\'\i}a-Bellido}, J. and {Gaztanaga}, E. and {Giannantonio}, T. and {Glazebrook}, K. and {Goldstein}, D.~A. and {Gonz{\'a}lez-Gait{\'a}n}, S. and {Gruen}, D. and {Gruendl}, R.~A. and {Gschwend}, J. and {Gupta}, R.~R. and {Gutierrez}, G. and {Hartley}, W.~G. and {Hinton}, S.~R. and {Hollowood}, D.~L. and {Honscheid}, K. and {Hoormann}, J.~K. and {Hoyle}, B. and {James}, D.~J. and {Jeltema}, T. and {Johnson}, M.~W.~G. and {Johnson}, M.~D. and {Kasai}, E. and {Kent}, S. and {Kessler}, R. and {Kim}, A.~G. and {Kirshner}, R.~P. and {Kovacs}, E. and {Krause}, E. and {Kron}, R. and {Kuehn}, K. and {Kuhlmann}, S. and {Kuropatkin}, N. and {Lahav}, O. and {Lasker}, J. and {Lewis}, G.~F. and {Li}, T.~S. and {Lidman}, C. and {Lima}, M. and {Lin}, H. and {Macaulay}, E. and {Maia}, M.~A.~G. and {Mandel}, K.~S. and {March}, M. and {Marriner}, J. and {Marshall}, J.~L. and {Martini}, P. and {Menanteau}, F. and {Miller}, C.~J. and {Miquel}, R. and {Miranda}, V. and {Mohr}, J.~J. and {Morganson}, E. and {Muthukrishna}, D. and {M{\"o}ller}, A. and {Neilsen}, E. and {Nichol}, R.~C. and {Nord}, B. and {Nugent}, P. and {Ogando}, R.~L.~C. and {Palmese}, A. and {Pan}, Y. -C. and {Plazas}, A.~A. and {Pursiainen}, M. and {Romer}, A.~K. and {Roodman}, A. and {Rozo}, E. and {Rykoff}, E.~S. and {Sako}, M. and {Sanchez}, E. and {Scarpine}, V. and {Schindler}, R. and {Schubnell}, M. and {Scolnic}, D. and {Serrano}, S. and {Sevilla-Noarbe}, I. and {Sharp}, R. and {Smith}, M. and {Soares-Santos}, M. and {Sobreira}, F. and {Sommer}, N.~E. and {Spinka}, H. and {Suchyta}, E. and {Sullivan}, M. and {Swann}, E. and {Tarle}, G. and {Thomas}, D. and {Thomas}, R.~C. and {Troxel}, M.~A. and {Tucker}, B.~E. and {Uddin}, S.~A. and {Walker}, A.~R. and {Wester}, W. and {Wiseman}, P. and {Wolf}, R.~C. and {Yanny}, B. and {Zhang}, B. and {Zhang}, Y. and {DES Collaboration}},
        title = "{First Cosmology Results using Type Ia Supernovae from the Dark Energy Survey: Constraints on Cosmological Parameters}",
      journal = {\apjl},
         year = 2019,
        month = feb,
       volume = {872},
       number = {2},
          eid = {L30},
        pages = {L30},
          doi = {10.3847/2041-8213/ab04fa},
archivePrefix = {arXiv},
       eprint = {1811.02374},
 primaryClass = {astro-ph.CO},
       adsurl = {https://ui.adsabs.harvard.edu/abs/2019ApJ...872L..30A}
}

@ARTICLE{PlanckCollaboration20,
       author = {{Planck Collaboration} and {Aghanim}, N. and {Akrami}, Y. and {Ashdown}, M. and {Aumont}, J. and {Baccigalupi}, C. and {Ballardini}, M. and {Banday}, A.~J. and {Barreiro}, R.~B. and {Bartolo}, N. and {Basak}, S. and {Battye}, R. and {Benabed}, K. and {Bernard}, J. -P. and {Bersanelli}, M. and {Bielewicz}, P. and {Bock}, J.~J. and {Bond}, J.~R. and {Borrill}, J. and {Bouchet}, F.~R. and {Boulanger}, F. and {Bucher}, M. and {Burigana}, C. and {Butler}, R.~C. and {Calabrese}, E. and {Cardoso}, J. -F. and {Carron}, J. and {Challinor}, A. and {Chiang}, H.~C. and {Chluba}, J. and {Colombo}, L.~P.~L. and {Combet}, C. and {Contreras}, D. and {Crill}, B.~P. and {Cuttaia}, F. and {de Bernardis}, P. and {de Zotti}, G. and {Delabrouille}, J. and {Delouis}, J. -M. and {Di Valentino}, E. and {Diego}, J.~M. and {Dor{\'e}}, O. and {Douspis}, M. and {Ducout}, A. and {Dupac}, X. and {Dusini}, S. and {Efstathiou}, G. and {Elsner}, F. and {En{\ss}lin}, T.~A. and {Eriksen}, H.~K. and {Fantaye}, Y. and {Farhang}, M. and {Fergusson}, J. and {Fernandez-Cobos}, R. and {Finelli}, F. and {Forastieri}, F. and {Frailis}, M. and {Fraisse}, A.~A. and {Franceschi}, E. and {Frolov}, A. and {Galeotta}, S. and {Galli}, S. and {Ganga}, K. and {G{\'e}nova-Santos}, R.~T. and {Gerbino}, M. and {Ghosh}, T. and {Gonz{\'a}lez-Nuevo}, J. and {G{\'o}rski}, K.~M. and {Gratton}, S. and {Gruppuso}, A. and {Gudmundsson}, J.~E. and {Hamann}, J. and {Handley}, W. and {Hansen}, F.~K. and {Herranz}, D. and {Hildebrandt}, S.~R. and {Hivon}, E. and {Huang}, Z. and {Jaffe}, A.~H. and {Jones}, W.~C. and {Karakci}, A. and {Keih{\"a}nen}, E. and {Keskitalo}, R. and {Kiiveri}, K. and {Kim}, J. and {Kisner}, T.~S. and {Knox}, L. and {Krachmalnicoff}, N. and {Kunz}, M. and {Kurki-Suonio}, H. and {Lagache}, G. and {Lamarre}, J. -M. and {Lasenby}, A. and {Lattanzi}, M. and {Lawrence}, C.~R. and {Le Jeune}, M. and {Lemos}, P. and {Lesgourgues}, J. and {Levrier}, F. and {Lewis}, A. and {Liguori}, M. and {Lilje}, P.~B. and {Lilley}, M. and {Lindholm}, V. and {L{\'o}pez-Caniego}, M. and {Lubin}, P.~M. and {Ma}, Y. -Z. and {Mac{\'\i}as-P{\'e}rez}, J.~F. and {Maggio}, G. and {Maino}, D. and {Mandolesi}, N. and {Mangilli}, A. and {Marcos-Caballero}, A. and {Maris}, M. and {Martin}, P.~G. and {Martinelli}, M. and {Mart{\'\i}nez-Gonz{\'a}lez}, E. and {Matarrese}, S. and {Mauri}, N. and {McEwen}, J.~D. and {Meinhold}, P.~R. and {Melchiorri}, A. and {Mennella}, A. and {Migliaccio}, M. and {Millea}, M. and {Mitra}, S. and {Miville-Desch{\^e}nes}, M. -A. and {Molinari}, D. and {Montier}, L. and {Morgante}, G. and {Moss}, A. and {Natoli}, P. and {N{\o}rgaard-Nielsen}, H.~U. and {Pagano}, L. and {Paoletti}, D. and {Partridge}, B. and {Patanchon}, G. and {Peiris}, H.~V. and {Perrotta}, F. and {Pettorino}, V. and {Piacentini}, F. and {Polastri}, L. and {Polenta}, G. and {Puget}, J. -L. and {Rachen}, J.~P. and {Reinecke}, M. and {Remazeilles}, M. and {Renzi}, A. and {Rocha}, G. and {Rosset}, C. and {Roudier}, G. and {Rubi{\~n}o-Mart{\'\i}n}, J.~A. and {Ruiz-Granados}, B. and {Salvati}, L. and {Sandri}, M. and {Savelainen}, M. and {Scott}, D. and {Shellard}, E.~P.~S. and {Sirignano}, C. and {Sirri}, G. and {Spencer}, L.~D. and {Sunyaev}, R. and {Suur-Uski}, A. -S. and {Tauber}, J.~A. and {Tavagnacco}, D. and {Tenti}, M. and {Toffolatti}, L. and {Tomasi}, M. and {Trombetti}, T. and {Valenziano}, L. and {Valiviita}, J. and {Van Tent}, B. and {Vibert}, L. and {Vielva}, P. and {Villa}, F. and {Vittorio}, N. and {Wandelt}, B.~D. and {Wehus}, I.~K. and {White}, M. and {White}, S.~D.~M. and {Zacchei}, A. and {Zonca}, A.},
        title = "{Planck 2018 results. VI. Cosmological parameters}",
      journal = {\aap},
         year = 2020,
        month = sep,
       volume = {641},
          eid = {A6},
        pages = {A6},
          doi = {10.1051/0004-6361/201833910},
archivePrefix = {arXiv},
       eprint = {1807.06209},
 primaryClass = {astro-ph.CO},
       adsurl = {https://ui.adsabs.harvard.edu/abs/2020A&A...641A...6P}
}

@ARTICLE{sotani_rev,
       author = {{Sotani}, Hajime},
        title = "{Understanding supernova gravitational waves with protoneutron star asteroseismology}",
      journal = {Classical and Quantum Gravity},
         year = 2026,
        month = may,
       volume = {43},
       number = {9},
          eid = {093001},
        pages = {093001},
          doi = {10.1088/1361-6382/ae6543},
       adsurl = {https://ui.adsabs.harvard.edu/abs/2026CQGra..43i3001S}
}

@ARTICLE{Sotani&Takiwaki20,
       author = {{Sotani}, Hajime and {Takiwaki}, Tomoya},
        title = "{Accuracy of the relativistic Cowling approximation in protoneutron star asteroseismology}",
      journal = {\prd},
         year = 2020,
        month = sep,
       volume = {102},
       number = {6},
          eid = {063025},
        pages = {063025},
          doi = {10.1103/PhysRevD.102.063025},
archivePrefix = {arXiv},
       eprint = {2009.05206},
 primaryClass = {astro-ph.HE},
       adsurl = {https://ui.adsabs.harvard.edu/abs/2020PhRvD.102f3025S}
}

@ARTICLE{Damour93,
       author = {{Damour}, Thibault and {Esposito-Farese}, Gilles},
        title = "{Nonperturbative strong-field effects in tensor-scalar theories of gravitation}",
      journal = {\prl},
         year = 1993,
        month = apr,
       volume = {70},
       number = {15},
        pages = {2220-2223},
          doi = {10.1103/PhysRevLett.70.2220},
       adsurl = {https://ui.adsabs.harvard.edu/abs/1993PhRvL..70.2220D}
}

@ARTICLE{Ramazanoglu16,
       author = {{Ramazano{\v{g}}lu}, Fethi M. and {Pretorius}, Frans},
        title = "{Spontaneous scalarization with massive fields}",
      journal = {\prd},
         year = 2016,
        month = mar,
       volume = {93},
       number = {6},
          eid = {064005},
        pages = {064005},
          doi = {10.1103/PhysRevD.93.064005},
archivePrefix = {arXiv},
       eprint = {1601.07475},
 primaryClass = {gr-qc},
       adsurl = {https://ui.adsabs.harvard.edu/abs/2016PhRvD..93f4005R}
}

@ARTICLE{Novak00,
       author = {{Novak}, J{\'e}r{\^o}me and {Ib{\'a}{\~n}ez}, Jos{\'e} M.},
        title = "{Gravitational Waves from the Collapse and Bounce of a Stellar Core in Tensor-Scalar Gravity}",
      journal = {\apj},
         year = 2000,
        month = apr,
       volume = {533},
       number = {1},
        pages = {392-405},
          doi = {10.1086/308627},
archivePrefix = {arXiv},
       eprint = {astro-ph/9911298},
 primaryClass = {astro-ph},
       adsurl = {https://ui.adsabs.harvard.edu/abs/2000ApJ...533..392N}
}

@ARTICLE{Murphy25,
       author = {{Murphy}, R. Daniel and {Mezzacappa}, Anthony and {Lentz}, Eric J. and {Marronetti}, Pedro},
        title = "{Core collapse supernova gravitational wave sourcing and characterization based on three-dimensional models}",
      journal = {\prd},
         year = 2025,
        month = sep,
       volume = {112},
       number = {6},
          eid = {063062},
        pages = {063062},
          doi = {10.1103/5z9g-yr28},
archivePrefix = {arXiv},
       eprint = {2503.06406},
 primaryClass = {astro-ph.HE},
       adsurl = {https://ui.adsabs.harvard.edu/abs/2025PhRvD.112f3062M}
}

@ARTICLE{Gerosa16,
       author = {{Gerosa}, Davide and {Sperhake}, Ulrich and {Ott}, Christian D.},
        title = "{Numerical simulations of stellar collapse in scalar-tensor theories of gravity}",
      journal = {Classical and Quantum Gravity},
         year = 2016,
        month = jul,
       volume = {33},
       number = {13},
          eid = {135002},
        pages = {135002},
          doi = {10.1088/0264-9381/33/13/135002},
archivePrefix = {arXiv},
       eprint = {1602.06952},
 primaryClass = {gr-qc},
       adsurl = {https://ui.adsabs.harvard.edu/abs/2016CQGra..33m5002G}
}

\end{document}